\documentclass[useAMS,usenatbib,usegraphicx]{mn2e}

\usepackage{times}
\usepackage{amssymb}
\usepackage{longtable}
\usepackage{subfig}
\usepackage[varg]{txfonts}
\usepackage{graphicx}
\usepackage{natbib}
\usepackage{longtable}
\bibpunct{(}{)}{;}{a}{}{,}

\newcommand{\kms}{$\rm km\,s^{-1}$}
\newcommand{\am}{$^{\prime}$}

\newcommand{\hi}{\mbox{H\,\sc{i}}}
\newcommand{\hii}{\mbox{H\,\sc{ii}}}

\title[The infrared and molecular environment surrounding the Wolf-Rayet star WR\,130]{The infrared and molecular environment surrounding  WR\,130}

\author[S. Cichowolski et al.] {S. Cichowolski$^1$\thanks{Member of
    the Carrera del Investigador Cient\'{\i}fico of CONICET,
    Argentina.}, L.A. Suad$^2$\thanks{Post-Doc Fellow of CONICET, Argentina}, S. Pineault$^3$, A. Noriega-Crespo$^4$, E.M. Arnal$^{2,5 *}$, and  N. Flagey$^6$\\
$^{1}$ Instituto de Astronom\'{\i}a y F\'{\i}sica del Espacio (IAFE), CC 67, Suc. 28, 1428 Buenos Aires, Argentina\\
$^{2}$ Instituto Argentino de Radioastronom\'{\i}a  (CCT-La Plata, CONICET), CC 5, 1894 Villa Elisa,
  Argentina\\
$^{3}$ D\'epartement de physique, de g\'enie physique et d'optique, Universit\'e Laval, Qu\'ebec, G1V 0A6 Canada, and Centre de recherche en astrophysique du Qu\'ebec (CRAQ)\\
$^{4}$ Space Telescope Science Institute, 3700 San Martin Drive, Baltimore, MD 21218, USA.\\
$^{5}$ Facultad de Ciencias  Astron\'omicas y Geof\'{\i}sicas, Universidad Nacional de La Plata, Paseo del
  Bosque s/n, 1900 La Plata, Argentina\\
$^{6}$  Canada-France-Hawaii Telescope Corporation, 65-1238 Mamalahoa Hwy, Kamuela, HI 96743, USA.}

\begin{document}

\date{}

\pagerange{\pageref{firstpage}--\pageref{lastpage}} \pubyear{}

\maketitle

\begin{abstract} We present a study of the molecular CO gas and mid/far infrared radiation arising from the environment surrounding
the Wolf-Rayet (W-R) star 130. We use the multi-wavelength data to analyze the properties of the dense gas and dust, and its
possible spatial correlation with that of  Young Stellar Objects (YSOs).
We use CO J=1-0 data from the FCRAO survey as tracer of the molecular gas, and mid/far infrared data
from the recent WISE and Herschel space surveys to study the dust continuum radiation and to identify a population of  associated candidate YSOs. 
The spatial distribution of the molecular gas shows a ring-like structure very similar to that observed in the \hi\, gas, and over the
same velocity interval.  The relative spatial distribution of the \hi\, and CO components is consistent with a photo-dissociation region. We have identified and characterized four main and distinct molecular clouds that create this structure.
 Cold dust is coincident with the dense gas shown in the  CO measurements.
We have found  several cYSOs that lie along the regions with the highest gas column density, and suggest that they are spatially
correlated with the shell. These are indicative of regions of star formation induced by the
     strong wind and ionization of the WR star.
\end{abstract}

\label{firstpage}


\section{Introduction}

Massive  O-type stars and their descendants  have a major impact on their surrounding interstellar medium 
(ISM) via their high throughput of ionizing photons, their energetic
 winds,  their expanding \hii\, regions, and their final explosion as  supernova.

 During their evolutionary courses, massive stars possess quite different mass loss rates and wind speeds,  generating  a variety of structures such as bubbles, shells (or multi-shells) in the ISM \citep{cas75, wea77, gar95}.
 Due to the high number of ultraviolet photons emitted by the star during the O-type phase, an \hii\, region is formed around the star \citep{wea77}. As this hot region expands, neutral material accumulates
 between the ionization front  and the shock front  that precedes it in
 the neutral gas \citep[see][]{dys97}. Simultaneously with the ionization of its 
surroundings, stellar winds are  blown by massive stars, which may  affect the evolution of the \hii\, region.

The following phase in the evolution of the star depends on its initial  mass  \citep[e.g.][]{cro07}. For masses higher than $\sim 60$ M$_{\odot}$ the star could go directly to a Wolf-Rayet (WR) phase, without going first throught a  RSG or LBV phase \citep[e.g.][]{smi08}.

During the WR phase, the stellar
 wind strengthens (could be an order of magnitude higher than in the main sequence phase) and may clear the region around the ionizing star, creating a stellar-wind bubble filled with high temperature (T $\sim$ 10$^{6-7}$ K) and
low density (n $\sim$ 10$^{-2}$ -- $10^{-3}$ cm$^{-3}$) shocked gas. In due time,
 this hot cavity may drive expanding gas shells that may eventually interact
 with the shell related to the expanding \hii\, region.
In this way, the supersonic winds can entrain and accelerate the surrounding gas, thereby injecting momentum and energy into the surrounding gaseous environment, strongly affecting its evolution.

The structures created by the action of  massive stars, usually striking at radio frequencies, are also very  bright at far infrared wavelengths, due to the presence of dust grains that absorb the UV stellar photons and cool down emitting in the infrared. 
The so-called infrared dust bubbles can now been studied in detail since the realease of the WISE \citep{wri10}, Spitzer \citep{ben03,car09}, and Herschel \citep{mol10} data \citep[e.g.][]{zav10,pal12}.

One important consequence of the interaction between masive stars and their local ISM is that the physical conditions present in the dense expanding shells are ideal for star formation activity.
Several studies in fact show the presence of young stellar object candidates ({\rm cYSOs})
whose birth was probably triggered by the expansion of  \hii\, regions \citep[and references there in]{zav10}.
However, similar studies towards Wolf-Rayet stars, representative of more
 evolved stages of massive stars, are scanty \citep{cap10,liu12, liu13}.
Thus, it is still unclear whether these older  bubbles   can create regions of active
star formation around them.

In this work we present the second part of a multiwavelength study of the surroundings of WR\,130, a WN8(h) star \citep{van01}  or, as proposed by \citet{smi08}, a WNH. 
The place that  this type of WR stars, which are the least chemically evolved \citep{con83}, has in the  evolution sequence of a massive star is not completely understood. 
Based on the mass discrepancy observed between WNH and H-poor WR stars (the WNH stars have higher masses), and on the fact that the most luminous WNH stars tend to be more H-rich than LBVs,
\citet{smi08} concluded that the WNH stars are   pre-LBVs, and that via  the violent eruptions of the LBV phase the star removes the hydrogen envelope and becomes a WN star. In this scenario, WR\,130 could be in a pre-LBV phase. 
The observed impact that WR\,130 has on its environment seems to support this possibility. Indeed,
a very well defined ring-like \hii\, region, first noticed by
\citet{hec82}, is associated with this star. Later, \citet{cic01} studied the region using a multi-wavelength approach that involved  {\rm DRAO}
\hi\, line data and continuum at 408 and 1420 MHz, H110$\alpha$ {\rm VLA}
 data, {\rm HIRES} 60 and 100 $\mu$m data \citep{fow94}, and 12 
and 25 $\mu$m Mid-Infrared Galaxy Atlas images \citep{ker00}. 
From  all the data analyzed, \citet{cic01} observed a well defined ring-like structure, which is consistent with a scenario were the action of the O star was directly followed by the action of the WR star, since the material expected to be present if the LBV phase took place is not detected.

In this paper we now present a study of the CO clouds  detected in the area of WR\,130 and analyze the properties of the dust making use of the new available WISE and Herschel data, which allow us to provide new insights into the gas and dust properties of the interstellar medium surrounding a Wolf-Rayet star. We also search for the presence of triggered star formation towards the surroundings of  WR\,130.

\section{Observations}

In this paper, we use the 1420 MHz radio continuum image of the WR\,130 area
from the Canadian Galactic Plane Survey \citep[CGPS -- ][]{tay03}.
Because this image is the result of a mosaic containing more than
one pointing field, it has a slightly better signal-to-noise
ratio and dynamic range.  It also has better flux and position
 registration, and a slightly better resolution, factors which are important when estimating the fluxes of some discrete sources
seen projected over an inhomogeneous background.

At infrared wavelengths, we have used the  data  retrieved from the 
Wide-Field Infrared Survey Explorer (WISE) \citep{wri10}, which is
an all-sky survey that contains images in four infrared channels centred at
3.4, 4.6, 12, and 22 $\mu$m. These channels, respectively, have
angular resolutions of 6.1, 6.4, 6.5, and 12 arcsec. We obtained all the images  from the NASA/IPAC Infrared Science Archive (http://irsa.ipac.caltech.edu).

We have also included {\it Herschel} data \citep{pil10}) from the $Herschel$ Infrared Galactic Plane Survey \citep[HiGAL,][]{mol10} obtained in its five photometric bands  at 70 and 160 $\mu$m with PACS \citep{pog10}, plus 250, 350, and 500 $\mu$m with SPIRE \citep{gri10}.   The maps have been produced by the improved version of
the ROMAGAL pipeline using UNIMAP, and therefore, are optimized to preserve the extended/diffuse emission, plus to avoid artifacts around bright point sources  \citep[see e.g.][]{tra11, pia15}.
The HiGAL maps have angular beams with a FWHM of about 5, 12, 18, 25 \& 36\arcsec\, at 70, 160, 250, 350 \& 500 $\mu$m, respectively. And the maps themselves were created with pixel sizes of 3.2, 3.2, 4.5, 6.0, 8.0, and 11.5\arcsec\, at 70, 160, 250, 350 \& 500$\mu$m, respectively.

Molecular CO $J= 1-0$ observations were obtained from the Five College Radio Astronomical Observatory
(FCRAO) CO Survey of the Outer Galaxy \citep{hey98}.
The angular resolution is 45\arcsec\,, the velocity resolution 0.127 \kms\,, and  the rms noise 0.2 K

\section{CO emission}\label{co}

\begin{figure*}
\centering
\includegraphics[width=17cm]{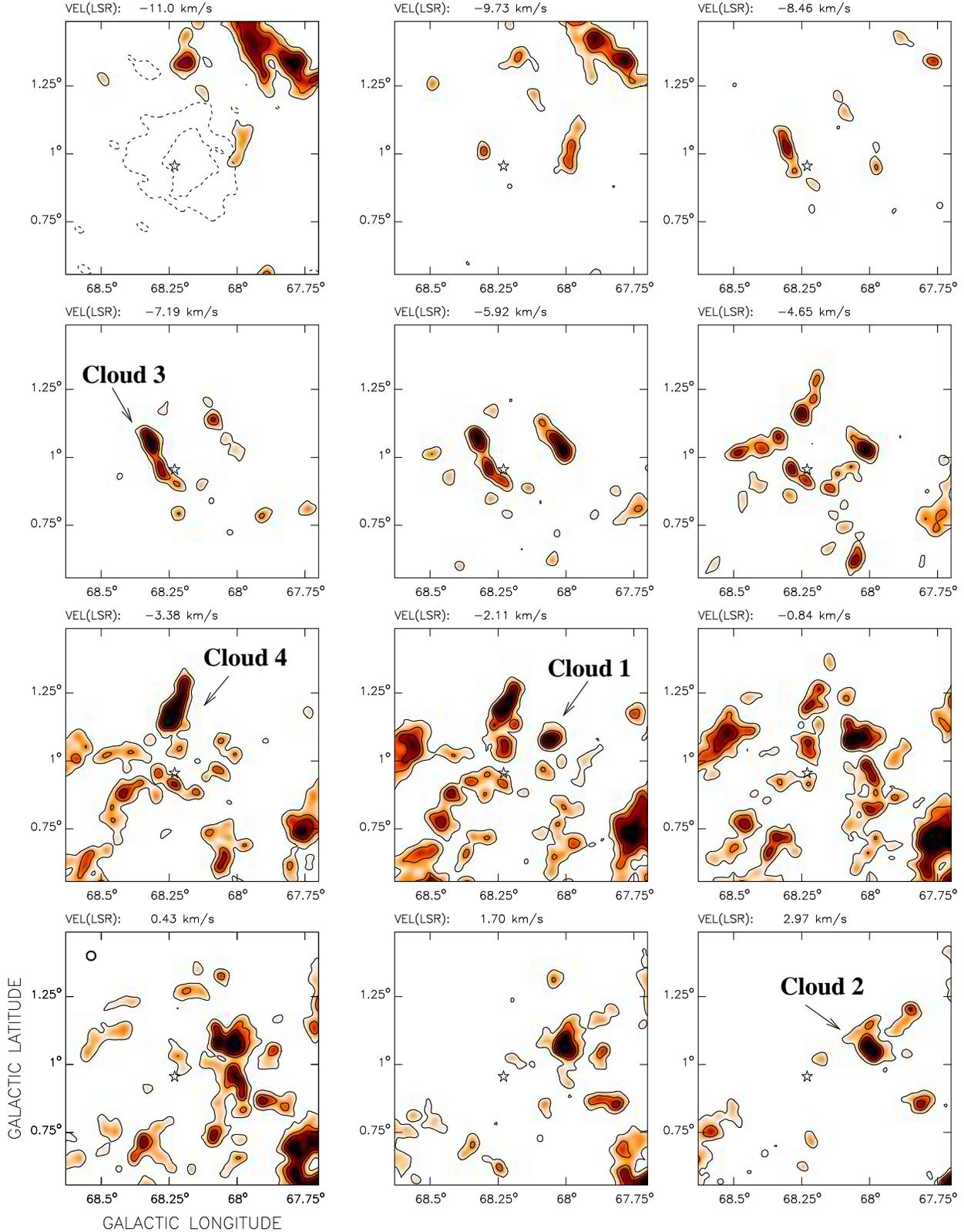}
\caption{CO emission distribution in the velocity range from --11 to 3 \kms. Each image is an average of ten consecutive velocity channels, where the corresponding central velocities are indicated in the top side of each box.  Contour levels are at 0.3, 0.7 and 1.1 K. Angular resolution is 2\am.  Beam shape is shown in the upper left corner of the bottom left panel. The 9.5 K level at 1420 MHz is indicated in the top left panel to delineate the ionized region found to be related to WR\,130 by \citet{cic01}. The star symbol indicates the location of WR\,130.}
\label{set-co}
\end{figure*}

Figure \ref{set-co} shows a set of images of the CO(1-0) emission distribution within the velocity range from about --11 to 3 \kms\, (all velocities
are with respect to the Local Standard of Rest (LSR)). 
Each image is an average of the emission in ten consecutive velocity channels, where the corresponding central velocities are indicated in the top side of each box. 
The velocity range shown was chosen according to the work of \citet{cic01}, where they found an \hi\, structure related to WR\,130 in the velocity range from --10.3 to 1.2 \kms.
To indicate the location of the \hii\, region found by \citet{cic01}, a 1420 MHz contour delineating it is shown in the top left panel.

\begin{figure*}
\centering
\includegraphics[width=17cm]{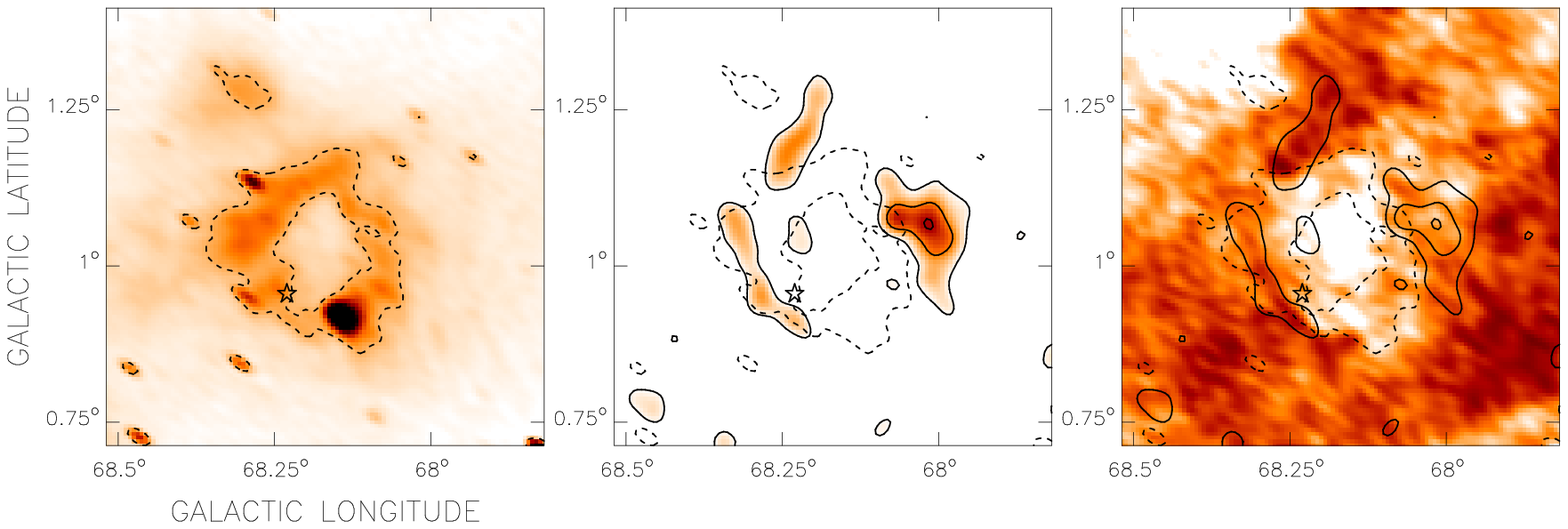}
\caption{{\it Left panel}: CGPS 1420 MHz emission distribution as shown in \citet{cic01}. Contour level is at 9.5 K. {\it Middle panel}: CO emission averaged between --11 and 3 \kms. Contours levels are at 0.3, 0.7 and 1.1 K. The dotted contour corresponds to the  9.5 K emission at 1420 MHz. {\it Right panel}: CGPS \hi\, emission distribution averaged between --10.3 and 1.2 \kms (as in \citet{cic01}) with the 1420 MHz and CO contours superimposed. In all the panels the star symbol indicates the location of WR\,130.}
\label{comp}
\end{figure*}

As can be seen in Fig. \ref{set-co} several molecular clouds are present in the area of WR\,130.
In particular, at --4.65 \kms\, the star lies, in projection, inside a well defined molecular shell.
To compare the molecular distribution with the ionized and atomic gas related to WR\,130 \citep{cic01}, in Fig. \ref{comp} the CO emission averaged between --11 and 3 \kms\, is shown, along with the 1420 MHz and \hi\, emissions.
The left panel of the figure shows the emission distribution  at 1420 MHz. The radio continuum emission in the area of WR\,130 was extensively analyzed by \citet{cic01}, who concluded that the ring like feature, named G68.1.1+1.1, is the radio counterpart of the optical region Sh2-98 and it is related to WR\,130. On the contrary, based on RRL data, \citet{cic01} found that the brightest compact source, named G68.14+0.92, observed at ($l, b$) = (68\fdg14, 0\fdg92) is an \hii\, region located at a distance of about 12 kpc, and thus unrelated to WR\,130.  

In the middle and right panels of Fig. \ref{comp}, the CO and \hi\, emissions are shown, respectively, with the radio continuum level delineating the \hii\, region superimposed.
From these images it is clear that the ionized gas is surrounded by an atomic shell and that several molecular clouds, which have probably formed part of the molecular cloud where WR\,130 was born, are still present.  The observed emission distributions clearly indicate that a photo-dissociated region (PDR) was formed around WR\,130.

To estimate the molecular mass of the clouds probably related to WR\,130 we first estimate the column density through the empirical relation

$$ N(H_{2}) = X \times I_{CO}$$
  
\noindent where $X$ is the CO-to-H$_2$ conversion factor, for which we considered the varying relation with the  Galactocentric radius R given by \citet{nak06},

$$ X [\rm cm^{-2}\,\rm K^{-1}\,\rm km\,\rm s^{-1}] = 1.4 \times 10^{20}\,  exp(R/11\, \rm kpc), $$

\noindent and $I_{CO}$ is the integrated line emission in the velocity range where the cloud is observed,
$I_{CO} = \int T\, dv$.

Figure \ref{comp} shows a comparison of the emission observed in the radio continuum at 1420 MHz, the CO and \hi\, emission, the latter two being averages over a few velocity channels.		
Assuming that the molecular gas shown in Fig. \ref{comp} (middle panel) is related to the \hi\, feature found by \citet{cic01} as being created by WR\,130, we adopt for the molecular clouds the distance inferred by \citet{cic01}, i.e. D = 5 kpc. We thus obtain that the structure is located at R = 8 kpc, which yields $X = 2.9 \times 10^{20}\, \rm cm^{-2}\,\rm K^{-1}\,\rm km\,\rm s^{-1}$.
Then, we estimate the total molecular mass related to WR\,130 based on the averaged emission shown in Fig. \ref{comp}, as well as the mass of the four individual molecular clouds  indicated in Fig. \ref{set-co}. These clouds were selected because they host infrared source candidates to be young stellar objects, as will be discussed in Section \ref{sf}. 
The molecular masses were derived from $M_{\odot} = 4.2 \times 10^{-20}\, N(H_{2})\, D^2\, A$, where D is the distance in pc and A is the solid angle in steradians.
Assuming an elliptical geometry for the individual clouds , the volume density can be estimated as $n(H_2)$ (cm$^{-3}$) = $10\,  M_{\odot} / a\, b^2$, where $a$ and $b$ are the major and minor semiaxes, respectively, in pc.
The masses and densities obtained, along with the velocity intervals considered for each estimate are shown in Table \ref{masas}.

We have also estimated the dust masses associated with each cloud, which are given in the last column of Table \ref{masas}. They were obtained using the Herschel data. For each cloud fluxes were measured at 70, 160, 250, 350, and 500 $\mu$m by rebinning all the images to the CO resolution and subtracting emission from the local background. 
Then  the spectral energy distribution
(SED) of each region was fitted using the models created by DUSTEM (see Section \ref{ir}) allowing the estimation of the dust masses associated with each region.

The kinetic energy stored in the CO shell can be estimated as $E_{\rm kin} = 0.5\, M_{\rm shell}\, V^2_{\rm exp}$, where $V_{\rm exp}$ is the expansion velocity of the shell and $M_{\rm shell}$ is the total (molecular, atomic and ionized) shell mass. 
Adopting an expansion velocity equal to half the velocity interval 
where the structure is observed, $V_{\rm exp} = 7.0 \pm 1.3$ \kms\,,  the molecular mass given in Table \ref{masas} and the atomic and ionized masses estimated by  \citet{cic01},  1450 and 3000 $M_{\odot}$, respectively,  we obtain 
$E_{\rm kin} =$ (2.5 $\pm$ 1.0) $\times 10^{49}$ erg, assuming a 40\% error  for the masses.
 Although \citet{cic01} concluded  
that WR\,130 could have alone created the observed structure, it is important to note that they did not take into account the molecular mass present in the shell, which considerably increases the kinetic shell energy. Thus, we can compare now the new value obtained for $E_{\rm kin}$ with the mechanical energy deposited in the ISM by the wind of the Wolf-Rayet star, $E_w = (0.7-2.2) \times 10^{50}$ erg \citep{cic01}.
We obtain $\epsilon = E_{\rm kin}$/$E_w = 0.007 - 0.5$. The ratio $\epsilon$ measures the energy conversion efficiency in the shell, and according to evolutionary models $\epsilon \le 0.2$ \citep{koo92}. Thus, not all the 
possible values of $\epsilon$ are compatible with the scenario where  the energy injected during  the Wolf-Rayet phase is enough to create the structure. In this case, the contribution of the energy injected during the O star phase and/or other massive stars, should be considered. 
 As mentioned in the Introduction, WR\,130 is a WNH star, and according to \citet{smi08} its age would be of about 2-3 Myr and its initial mass of at least 60 M$_{\odot}$. A rough estimation of the energy injected by such a star during its main sequence yields $E_w = (2.5 - 3.5) \times 10^{50}$ erg \citep{jag88}, which would be enough to create the observed structure.
We have nevertheless looked for the presence of other massive stars in the region. 
We queried the available catalogues such as the  Galactic O-Star Catalog  \citep{mai13}, the
Early-Type Emission-Line Stars Catalogue \citep{wac70}, the 
Catalogue of Be stars \citep{jas82},  the 
H-alpha Stars in the Northern Milky Way Catalogue \citep{koh97}, and the Catalog of Galactic OB Stars \citep{ree03}, for early-type and emission stars. 
No stars were found in any catalogue. 
The only massive star located near by is, as mentioned by \citet{cic01}, an OB star, which has an   uncertain spectral type and no distance estimate \citep{sto60}. 
It is located  in projection not in the center of the structure but onto the shell (there is a second OB star mentioned by \citet{cic01}  but its location is actually outside the structure, see Figure 1 of \citet{cic01}).
Although we can not completely rule out the possibility that the OB star may be playing a role in creating the shell structure, we think that the action of WR\,130 is sufficient and most likely dominant in the region.

\begin{table}
\caption{Physical parameters of the molecular gas.}
\label{masas}
\centering
\begin{tabular}{l c c c c c}
\hline\hline
& $v_1$ & $v_2$ & Mass & Density &  Dust Mass\\
& (\kms) & (\kms) & ($10^3 M_{\odot}$) & ($10^3$ cm$^{-3}$) & $M_{\odot}$\\
\hline
Total & --11 & 3 & 46 & -- &--\\
Cloud 1 & --3.5& 1.84 & 2.2 & 1.2 & 114\\
Cloud 2 & --2.35& 4.25 & 4.6 & 2.4 & 180\\
Cloud 3 & --10.5 & --5.0 & 2.6 & 0.8 & 125\\
Cloud 4 & --7.18 & --0.7 & 6.0 &0.8 & 252\\
\hline
\end{tabular}
\end{table}

\section{Infrared emission  and dust properties }\label{ir}

In this section we make use of the WISE
and Herschel  data to analyze the distribution of the dust as well as its properties.

\begin{figure}
\centering
\includegraphics[width=9cm]{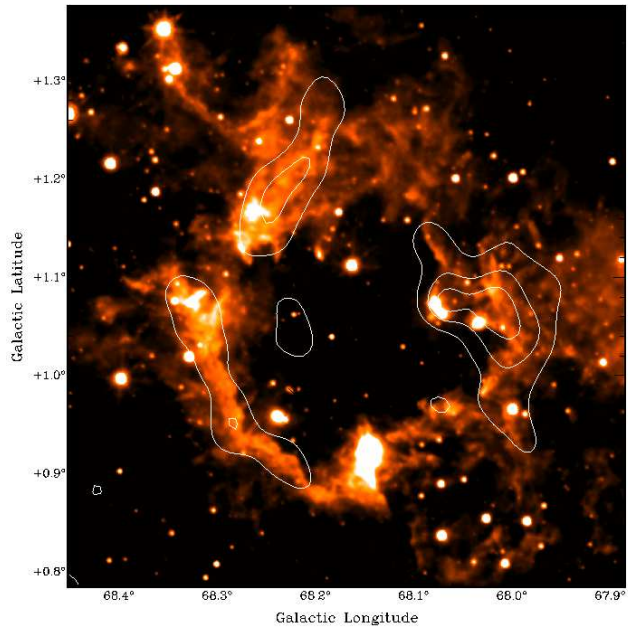}
\caption{WISE 12 $\mu$m image of the region. The contour levels correspond to the averaged CO emission between --11 and 3 \kms\, and are at 0.3, 0.6, and 0.9 K. }
\label{12mic}
\end{figure}

As shown in Section \ref{co}, the ionized region related to WR\,130 is interacting with the molecular gas. As a  consequence a PDR is created in the interior layers of the molecular clouds, which can be seen as polycyclic aromatic hydrocarbon (PAH) emission in WISE A  at 12 $\mu$m.  
Figure \ref{12mic} shows that the 12 $\mu$m emission displays a cavity surrounded by a roughly thin annular shell emission.

\begin{figure}
\centering
\includegraphics[width=9cm]{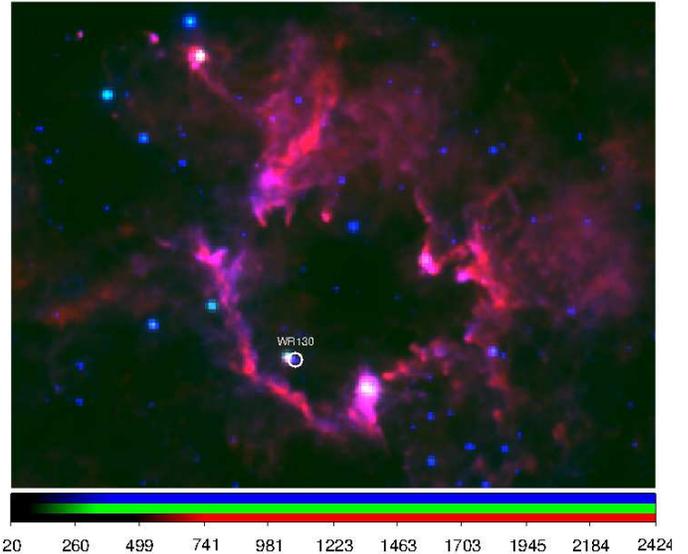}
\caption{Color composite image of the area of WR\,130. Blue color shows the emission at 12 $\mu$m (W3 WISE band), green color represents the emission at 22 $\mu$m (W4 WISE band), and red color shows the emission at 250 $\mu$m (SPIRE). The location of WR\,130 is indicated by the white circle.}
\label{wh}
\end{figure}

Figure \ref{wh} shows the distribution of the warm and cold dust traced by the 12 $\mu$m W3 band (PAHs; blue), 22 $\mu$m W4 band (VSGs; green) and Herschel 250 $\mu$m (Big Grains; red).
Overall the spatial distribution of the three components looks quite similar, at least at the angular resolution of the WISE data, 6 and 12 arcseconds for W3 and W4, respectively.
The biggest difference takes place on the rim near the WR\,130 source itself, where  a  warmer dust component appears closer to the source, suggesting a more
typical photodisociation structure, with the warm gas facing the ionizing source. WR\,130 itself, is detected by WISE at 12 $\mu$m (blue source inside the 1 arcmin radius
white circle).

\begin{figure*}
\centering
\includegraphics[width=14cm]{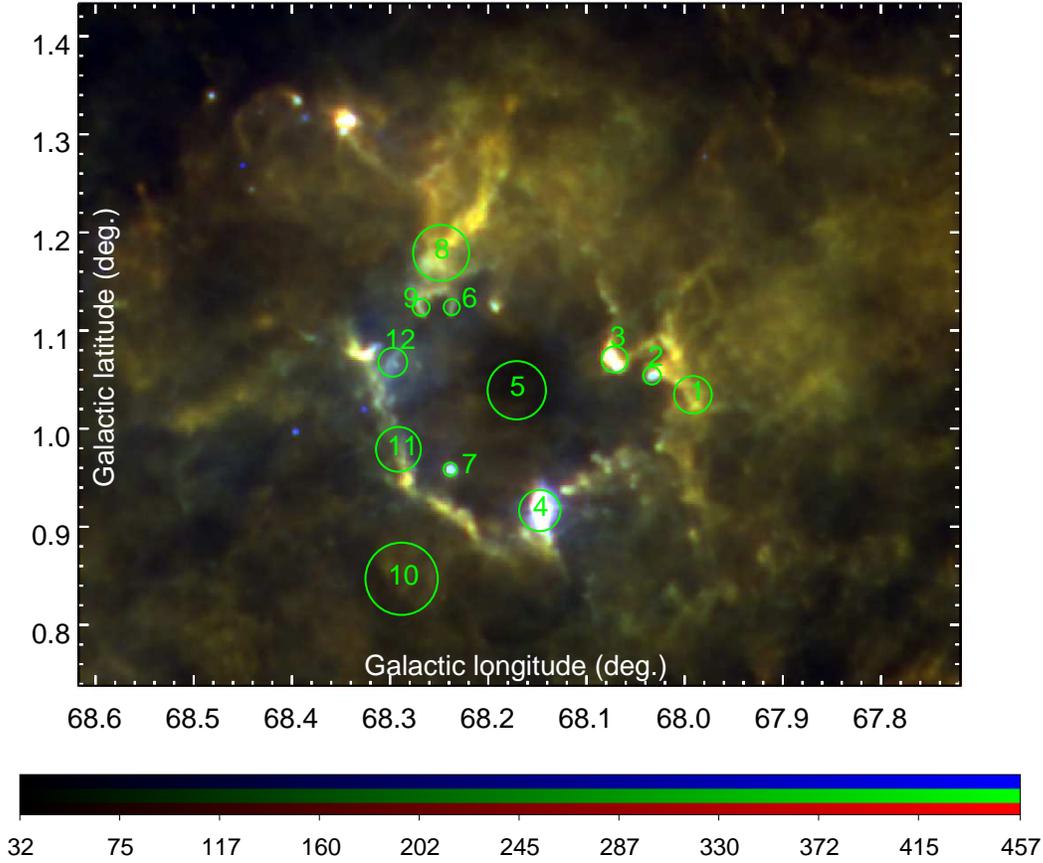}
\caption{Composite image of the region around WR\,130. The image shows SPIRE 500 $\mu$m emission in red, SPIRE 250 $\mu$m in green and PACS 70 $\mu$m in blue. Outlined are the regions where  flux densities were estimated. 
}
\label{rgb}
\end{figure*}

\begin{figure}
\centering
\includegraphics[width=9cm]{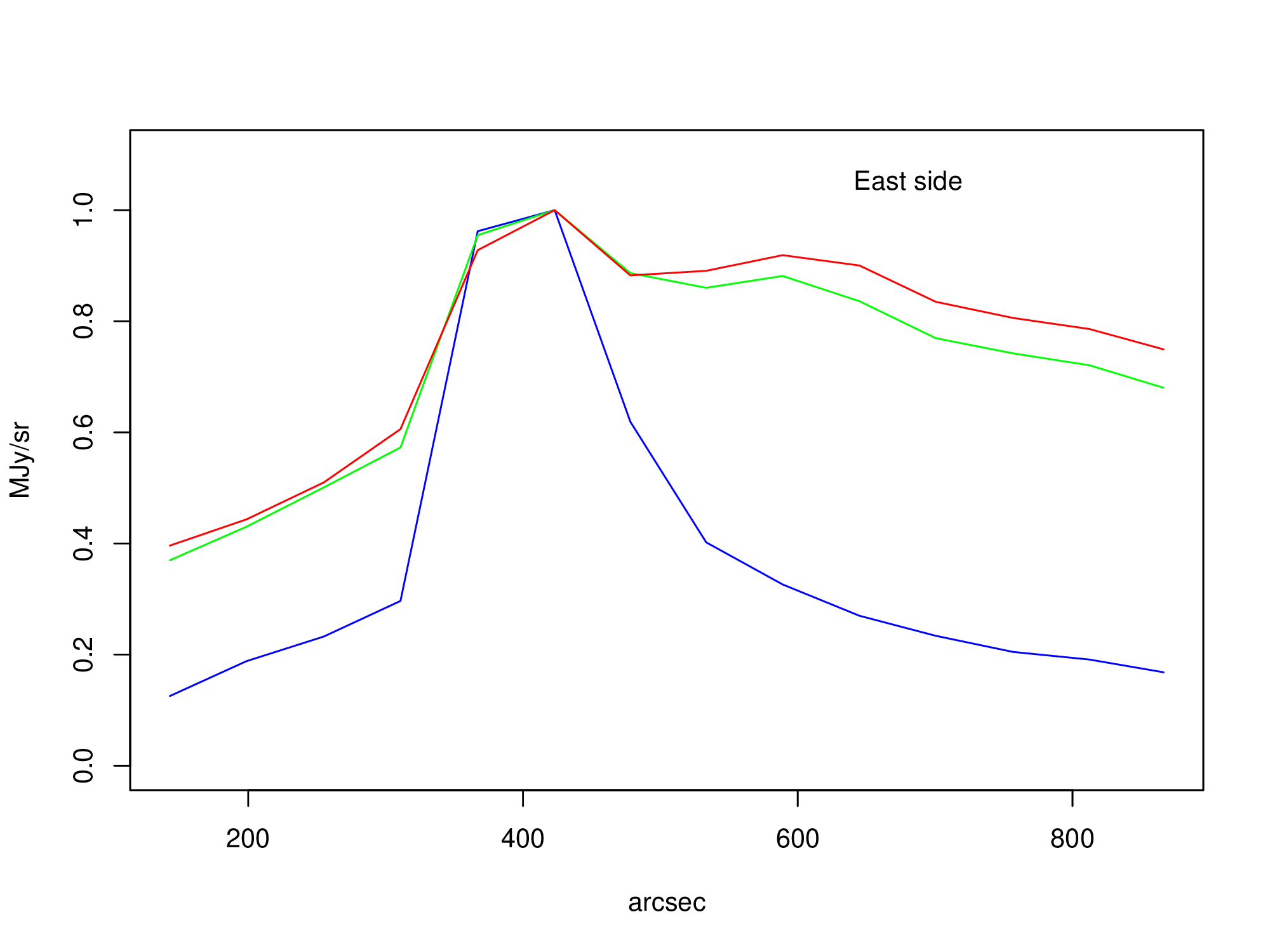}
\includegraphics[width=9cm]{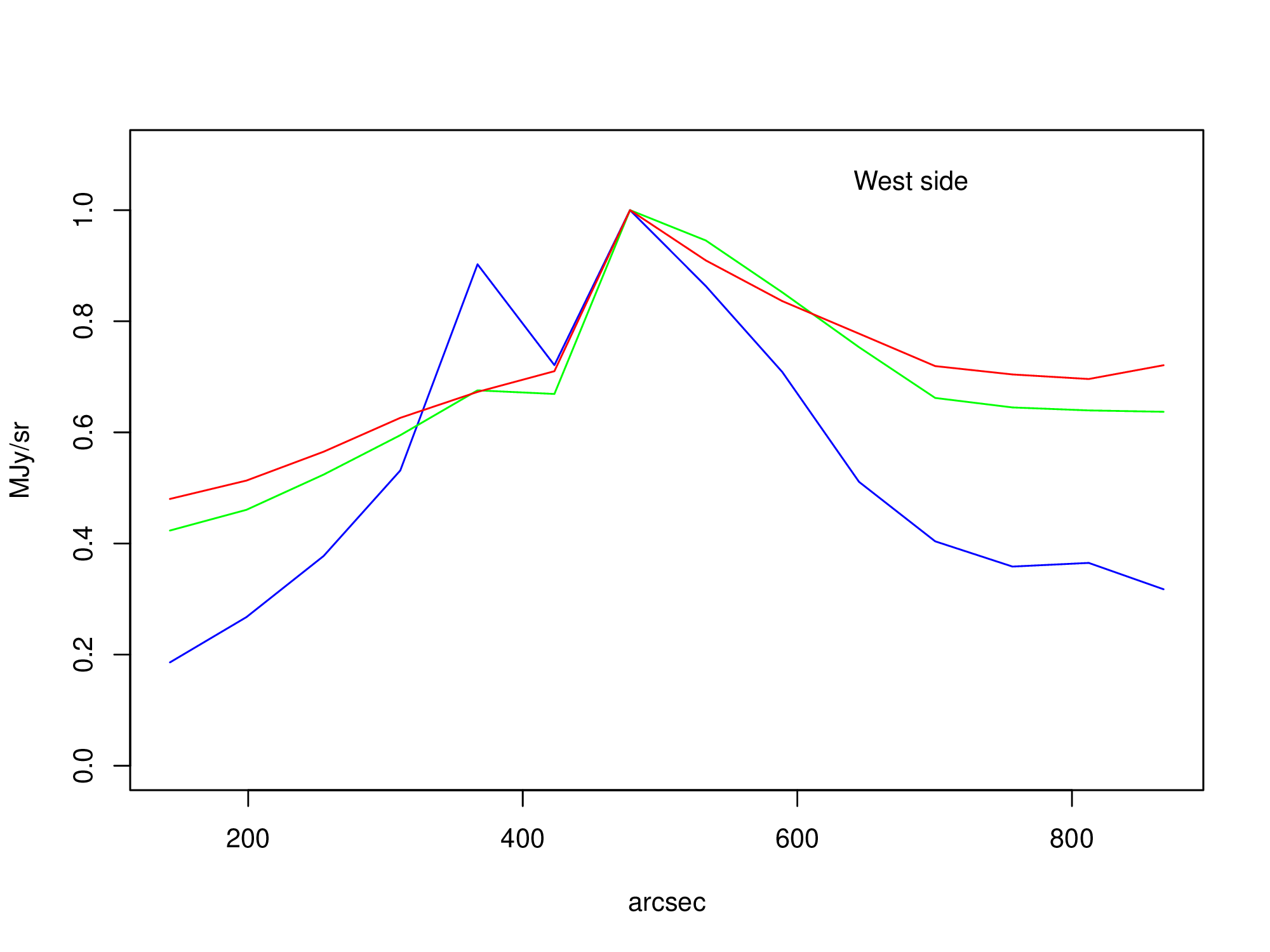}
\caption{Averaged 70 $\mu$m (blue), 250 $\mu$m (green), and 500 $\mu$m (red) infrared emission distributions as a function of angular distance from the center of the structure, ($l, b$) = (68\fdg17, 1\fdg035). {\it Top panel}: East side of the structure. {\it Bottom panel}: West side.}
\label{iring}
\end{figure}

The emission in  Herschel bands corresponding to 70 (blue), 250 (green) and 500 $\mu$m (red) is shown 
as a color-composite image in Fig. \ref{rgb}, where several regions of interest can be seen.
As can be noticed, the emission  observed in the three different  bands is similar, in the sense that it forms a shell-like structure, instead of having, as observed in several \hii\, regions, the emission at 70 $\mu$m brighter in the interior of the bubble, where the hotter material is expected to be present, 
 and the SPIRE emission, mapping the cooler dust, outside it \citep[e.g.][]{rod10, deh10, and12}.
In this context, the observed  large-scale distribution of the infrared emission at the periphery of the ionized gas represents matter accumulated during the expansion of the ionized gas.
Additional evidence supporting  this conclusion can be seen by comparing the emission distribution among the different infrared bands, as  shown in Fig \ref{iring}. In this figure the different  lines represent the mean value of the corresponding emission along concentric rings spaced by 1 arcmin and centred at ($l, b$) = (68\fdg17, 1\fdg035). We divided the structure in two halves, the West and  East sides, shown in the top and bottom panels of Fig. \ref{iring}, respectively.
The normalized emission was computed by dividing every point by the corresponding maximum value, 174.5 , 215.2 and 39.4 MJy/sr at 70, 250, and 500 $\mu$m respectively, for the East side; and 114.9, 223.1, and 40.2 MJy/sr for the West half. 
From this comparison it can be seen that the emissions in the three bands present a similar behavior and peak together at the location of the shell structure. 
It can also be noticed that in the West side the emission at 70 $\mu$m peaks twice. In this case the maximum  located closer to the center of the structure is caused by the strong emission observed bordering the molecular cloud labeled as Cloud 1 in Fig. \ref{set-co}.

The fact that the emission distributions show a similar structure at 70 $\mu$m and at longer wavelengths such as 500 $\mu$m is consistent with the evolutionary path suggested by \citet{smi08} for a WNH star, where, as mentioned in the Introduction of this paper, the energy injected by  the WR star impinges directly onto the bubble created during the O-type phase, making it possible that all the dust had been sweep-up into a shell structure.
On the other hand,  the spatial correlation between the infrared and CO emission is noticeable, suggesting that the infrared emission originated in  dense gas, where several CO, $^{13}$CO, CH+, and  NII transition lines can take place and be responsible  for most of the observed emission, as was detected in the area of the \hii\, region Sh2-104 \citep{rod10}.
As can be seen in Fig. \ref{12mic} the dense structure is a  fragmented shell of collected material (e.g. cloud 3) with few probably pre-existing clumps  (e.g. clouds 1 and 4) that clearly distorted the ionization front.

Given that  the physical conditions of the structure can be determined from the dust thermal emission, we have selected several regions (see Fig. \ref{rgb}) for which we obtained the  cold dust temperature using the estimated   Herschel fluxes, given in Table \ref{fluxes}. To get the flux density in each band, the background, chosen far from the structure, was first subtracted.
The regions were selected in order to sample dust related to different parts of the structure, i.e., the interior, the shell, the photo-dissociation region, the CHII region candidate and  cYSOs (see Section \ref{sf}), and the  \hii\, region G68.14+0.92 located at a much larger distance.

\begin{table*}
\caption{ Infrared flux densities estimated for regions 1 to 12.} 
\label{fluxes}
\centering  
\begin{tabular}{l c c c c c c c c c}
\hline \hline
\multicolumn{10}{c}{Flux Densities (Jy)} \\
Region & $l$(deg.) & $b$(deg.) & 12($\mu$m) & 22($\mu$m) & 70($\mu$m) & 160($\mu$m)& 250($\mu$m)& 350($\mu$m)& 500($\mu$m)\\
\hline 
1 & 67.99 & 1.03 & 0.4 & 0.7 & 6.4 & 32.5 & 26.3 & 13.4 & 5.7\\
2 & 68.03 & 1.05 & 0.6 & 0.9 & 13.8 & 20.6 & 11.5 & 4.4 & 2.2\\
3 & 68.07 & 1.07 & 0.9 & 2.2 & 35.7 & 55.3 & 32.1 & 16.0 & 6.3\\
4 & 68.15 & 0.92 & 4.3 & 16.1 & 191.3 & 212.1 & 95.6 & 36.2 & 15.1\\
5 & 68.17 & 1.04 & 0.2 & 0.3 & 0.6 & 5.8 & 5.7 & 2.9 & 1.4 \\
6 & 68.24 & 1.12 & 0.1 & 0.2 & 2.4 & 6.6 & 4.5  & 2.2 & 1.2 \\
7 & 68.24 & 0.96 & 0.7 & 5.5 & 40.9 & 23.5 & 8.8 & 3.3 & 1.2\\
8 & 68.25 & 1.18 & 1.3 & 1.8 &19.3   & 86.0  &66.0 &30.2 & 13.8\\
9 & 68.27 & 1.12 & 0.1 & 0.3 & 3.2 & 8.2 & 6.2 & 3.1 & 1.3\\
10 & 68.29 & 0.85 & 0.4 & 0.8 & 6.6 & 32.7 & 28.3 & 13.8 & 6.2\\
11 & 68.29 & 0.98 & 0.6 & 1.1 & 14.2 & 38.2 & 24.0 & 10.5 & 4.7\\
12 & 68.30 & 1.07 & 0.5 & 1.2 & 12.5 & 25.5 & 14.5 & 6.4 & 2.5 \\
\hline
\end{tabular}
\end{table*}

The spectral energy distributions (SEDs) obtained for the 12 regions, normalized to the 160 $\mu$m flux value, are shown in Fig. \ref{irseds}. An inspection of the distributions shows that all  regions but three have a similar SED and are  therefore expected to contain dust with similar properties. The regions that present  a different SED are indicated with different colors and correspond to region 4 (green line), region 5 (red line) and region 7 (orange line). From Figure \ref{rgb} it can be seen that  region 4 is associated with G68.14+0.92 (the farthest \hii\, region, which is clearly visible in the infrared), region 5 is located in the interior of the shell and region 7 corresponds to the CHII region candidate (see Section \ref{sf}.). 

To characterize the dust located in each region, the spectral energy distributions
(SEDs) were fitted using the models created by DUSTEM, 
and a full description of the assumptions and tests performed 
can be found in the original paper by \citet{com11}.
In a nutshell, DUSTEM creates spectral energy distributions of the 
emission from interstellar dust covering a wavelength range from the mid to far-IR
(approximately 3 to 1000 $\mu$m). The emitting dust is the result of three main components: polycyclic aromatic hydrocarbons (PAHs), amorphous carbon and 
amorphous silicates. The abundances and size distribution of each of these 
components and their interaction with the interstellar radiation field (ISRF), determine
essentially the shape of the SED. Roughly, PAHs are expected to dominate the spectra
between ~3-15 $\mu$m, small amorphous carbon grains the ~20-40 $\mu$m range and
larger grains (amorphous carbon and silicates) the longer wavelengths 
\citep[see e.g.][Fig.2]{com11}.
The present version of the code, although it is able
to deal with ISO and Spitzer mid-IR data, needs to be modified to include WISE
data. Therefore, in this study we are concentrating on the longer wavelength
range covered by the Herschel Space mission. In terms of determining a dust temperature
representative of the medium, the large grains are the ones that can reach
a thermal equilibrium with the ISRF, while smaller dust particles
are stochastically heated, and therefore, a dust equilibrium temperature
is meaningless.

We thus estimated the dust properties of four different regions, the  ones mentioned above which present a different SED (see Fig. \ref{irseds}) and region 11 as a reference for the rest of the regions. 
For each region the models  provide the ISRF needed to reach the thermal balance  \citep{mat83}, and
 the temperatures of the three dust components  (small carbon, large carbon and silicates). 
All the estimates  are shown in Table \ref{irprop}.

As can be seen in Table \ref{irprop} the estimated temperatures for the \hii\, region G68.14+0.92 and the candidate CHII (see Section \ref{sf}) are typical for \hii\, regions and higher than the ones found for the shell structure and its interior. On the other hand, the interior region is the coldest, with a temperature around 20 K, which is characteristic of the more diffuse interstellar
medium \citep{bou96} {and consistent with the lack of bubble material in this region.
As for the ISRF factor, it can be noted that it is also larger in regions 4 and 7, as expected for regions near a strong radiation field, as is the case for these regions, since they host new stars (see Section \ref{sf}). On the contrary, for the interior of the shell (region 5) and the shell itself (region 11) the model yields a very low ISRF, suggesting that  strong radiations fields are not involved in these regions.

\begin{figure}
\centering
\includegraphics[width=9cm]{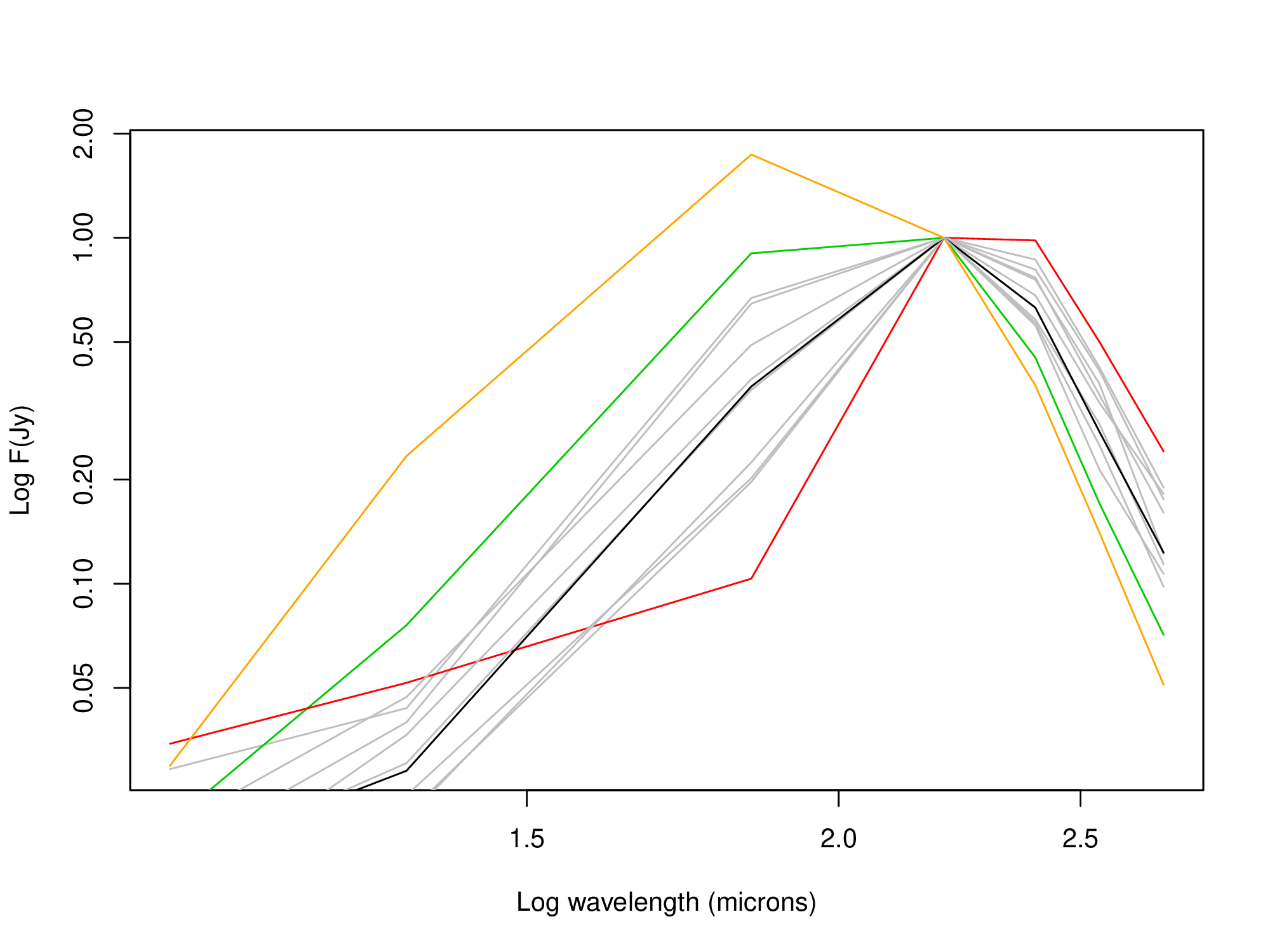}
\caption{Infrared flux densities measured for the 12 regions shown in Fig. \ref{rgb} at different wavelengths. The green, red, and  orange lines correspond to regions 4, 5 and 7, respectively (see Fig. \ref{rgb}). The   black line was taken as the representative infrared energy distribution of the other 9 regions  (shown in gray).}
\label{irseds}
\end{figure}

\begin{table*}
\caption{Dust properties of the four selected regions.} 
\label{irprop}
\vskip 0.25truecm
\centering  
\begin{tabular}{l  c c c  c}
\hline \hline
Region  &  $T_{\rm dust}$--small Carbons & $T_{\rm dust}$-- large Carbons   & $T_{\rm dust}$--silicates (K) &ISRF \\
\hline
4 (G68.14+0.92)& 34.4& 32.5& 22.3  & 8.95\\
5 (interior) & 20.1 & 19.0 &13.7 &  0.46\\
7 (CHII)& 41.8 & 39.4 &26.6 &  25.5\\
11 (shell)& 25.5 & 24.1 &17.0 &1.72\\
\hline
\end{tabular}
\end{table*}

\section{Young stellar population}\label{sf}

As shown in the previous sections, it is clear that the interstellar medium surrounding WR\,130 is strongly affected by the high UV radiation and mass loss rate of the star. 
As proposed by \citet{elm98}, in this scenario it is highly possible that  induced star formation  is taking place. With this in mind, we look for  primary tracers of stellar formation activity in the area making used of the IRAS Point Source Catalogue \citep{bei88}, the MSX Infrared Point Source Catalogue \citep{ega03}, the WISE  All-Sky Source Catalogue \citep{wri10},  and a preliminary source extraction from {\it Herschel} HiGAL data (Molinari et al. 2015, in preparation), using the CUTEX source extraction and photometry \citep{mol11}.
Within a circular area of 20\arcmin\, radius, centered at ($l, b$) = (68\fdg2, 1\fdg0), a total of 29 IRAS, 10 MSX (having flux quality $q > 1$ in the four bands),  686 WISE (with signal-to-noise ratios greater than 7 and photometry flux uncertainties lower than  0.2 magnitudes, in the W1, W2 and W3 bands), sources  were found.

To identify the young stellar object candidates (cYSOs) among the infrared sources, we adopted the  classification scheme described in \citet{jun92}, \citet{lum02}, and \citet{koe12}, for the IRAS, MSX and WISE data, respectively.  For the Herschel data we have 
used the prescription developed by the HiGAL team \citep{eli13} , where essentially a source is considered a candidate YSO if it is detected at 70 $\mu$m and appears in at least 3 contiguous bands.

In this way, we found that 2 IRAS sources, IRAS\,19563+3114 and IRAS\,19571+3113, are cYSOs. 
As for the MSX sources, according to the criteria of \citet{lum02}, 
we have identified massive young stellar object (MYSO) candidates 
and compact \hii\, region (C\hii) candidates. In this region, among the 10 MSX listed  sources we found 2 MYSO  and 3 C\hii\, candidates. 
We then analyzed the listed sources from the WISE catalogue and, following \citet{koe12}, 
before attempting to identify the cYSOs, we first selected the non-stellar  sources, such as PAH-emitting galaxies, broad-line active galactic nuclei (AGNs), unresolved knots of shock emission, and PAH-emission features.
A total of 119 sources were  dropped from the list, and among the remaining 567  only 2 were identified as Class I sources (e.g. sources where the IR emission arises mainly from a dense infalling envelope) and 9 as Class II sources (e.g. sources where the IR emission is dominated by the presence of an optically thick circumstellar disk).
In Fig. \ref{cc} we show the (W2 -- W3) vs (W1 -- W2) color-color diagram for all the uncontaminated sources, where the Class I and Class II sources are shown in red and green, respectively.
Given that protostellar objects with intermediate/high masses can be identified among the Class I sources by additionally requiring that the magnitude  W3 $<$ 5 \citep{hig13}, and that neither of the detected 2 Class I sources satisfies this criterion, it seems that both are low mass protostars.

 In the far-infrared, using the {\it Herschel} data, one is systematically detecting more deeply embedded objects; although is clear that those less embedded are likely to be detected at shorter wavelengths as well. Unlike low mass protostars where there is a general agreement on their classification as function of their SED and evolutionary stage, from Class 0 to Class III,
indicating their age and the clearance of their proto-stellar envelope  \citep[see e.g.][]{eva09}, for massive YSOs is not the case. It is possible, however, under the 'mass accretion paradigm' of massive star formation  \citep[see e.g.][and references therein]{tan14}, to develop a similar scheme where the youngest more embedded massive YSOs are identified as  
Class 0 and so on \citep{mol08}. Observationally, the HiGAL team has developed a method to select YSO 'candidates' that allows such classification and places the sources in a  bolometric luminosity versus mass envelope diagram,
providing an idea on their evolutionary stage.
The scheme, as mentioned above, requires a detection at 70 $\mu$m and three contiguous bands (to create a SED); for further details see \citep{eli13}.
In this study we have used such scheme, but we have gone a step further by trying to fit a SED model to data  when possible, and using the Robitaille-Whitney {\it Hyperion} grid of models for such task \citep{rob14}.

\begin{figure}
\centering
\includegraphics[width=9cm]{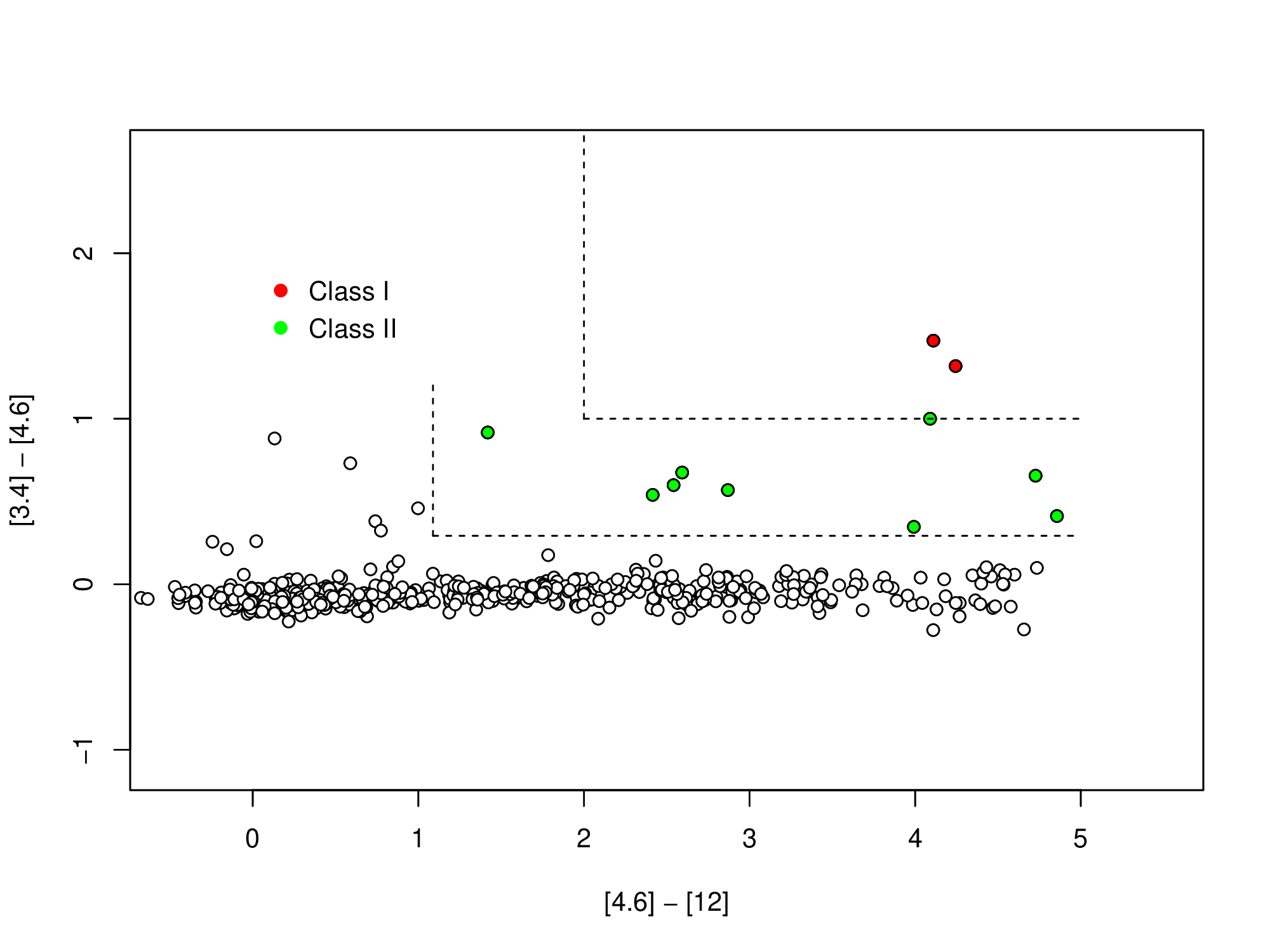}
\caption{WISE color-color diagram showing the areas where the Class I and Class II YSO candidates are located according to the \citet{koe12} criteria. }
\label{cc}
\end{figure}

\begin{figure}
\centering
\includegraphics[width=9cm]{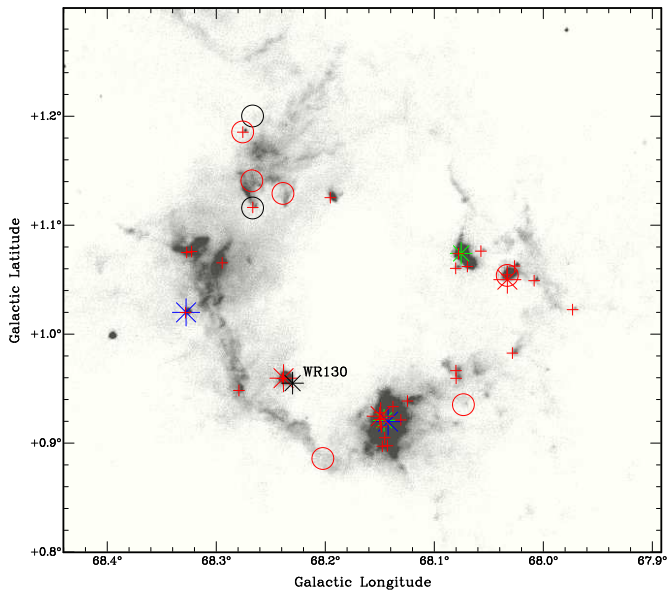}
\caption{Herschel image at 70 $\mu$m with the location of the cYSOs superimposed.  Green, blue, and red asterisks correspond to IRAS YSO, MYSO, and  C\hii\, candidates, respectively, while black and red  circles indicate the location of WISE Class I and Class II  candidate sources, respectively.  Herschel sources are indicated by red crosses. }
\label{figysos}
\end{figure}

\begin{table*}
\caption{ IRAS, MSX, WISE  and Herschel sources found to be YSO candidates.  \label{ysos}}
\begin{tabular}{c c c c c c c c}
\hline
\multicolumn{8}{c}{{\bf IRAS sources}} \\
\hline
 $\#$ &  Designation &  {(\it l, b})&  F$_{12}$[Jy] (Q$_{12}$) &  F$_{25}$[Jy] (Q$_{25}$) &  F$_{60}$[Jy] (Q$_{60}$) &  F$_{100}$[Jy] (Q$_{100}$) & Notes \\
\hline
1&19563+3114  & 68.07, 1.07 & 1.690 (3)& 4.411 (3)& 56.06 (3)&  105.2 (2)&  Cloud 1 \\
2 & 19571+3113 & 68.146, 0.92 & 3.682 (1: upper limit) & 15.050 (2)&  173.80  (3)& 399.9  (2) & G68.14+0.92  \\
\hline
\multicolumn{8}{c}{{\bf MSX sources}} \\
\hline
 $\#$ &  Designation &  {(\it l, b})&  F$_{8}$[Jy] (Q$_{8}$) &  F$_{12}$[Jy] (Q$_{12}$) &  F$_{14}$[Jy] (Q$_{14}$) &  F$_{21}$[Jy] (Q$_{21}$) & Notes \\
\hline
3&  G068.0331+01.0543 & 68.03, 1.05 &  0.74298 (4)&  1.3394 (1) &  0.85034 (1) &  2.7624 (1) & CHII, Cloud 2\\
4&  G068.1492+00.9247 &  68.149, 0.92 & 1.08850 (4)&  1.9132 (3) & 1.42180 (3) &  3.0316  (3) &CHII, G68.14+0.92\\
5& G068.2383+00.9596 & 68.238, 0.9596 & 0.96607(4) & 1.4840 (3) & 1.47640 (3) & 8.0493 (4) & CHII, CO also at --62 \kms \\
6& G068.3277+01.0201 & 68.32+01.02 & 0.60381 (4)& 2.30430 (4) &  2.8046 (4)&  4.9903 (4) &MYSO, Cloud 3\\
7& G068.1424+00.9197 & 68.14, 0.919 &  0.37255 (4) &  0.69779 (1) & 1.2673 (3)& 3.4386 (3) & MYSO, G68.14+0.92\\
\hline
\multicolumn{8}{c}{{\bf WISE sources}} \\
$\#$ &  Designation &  {(\it l, b})& W1 [mag]  &    W2 [mag]   &  W3 [mag]  &  W4 [mag] &Notes \\
\hline
8&J195819.33+313643.5 & 68.266, 1.20 & 14.366&  12.894 & 8.784 & 7.292& Class I, Cloud 4, CO also at --60 \kms\\
9&J195839.60+313404.4 & 68.266, 1.11 & 13.436 & 12.118 & 7.874 & 5.252 & Class I, Cloud 4\\
10& J195824.18+313643.7 & 68.27, 1.185 &  9.115 &  8.575 &6.160 & 3.996 & Class II, Cloud 4\\
11& J195833.72+313453.6& 68.267, 1.140& 11.840& 11.493 & 7.501 & 5.581 & Class II, Cloud 4\\
12&J195820.08+312011.3& 68.033, 1.053 &  9.388&  8.388 &4.298 & 2.370 & Class II, Cloud 2\\
13& J195925.21+312333.1 & 68.20, 0.88 & 14.019 & 13.607 & 8.751 & 7.244 & Class II, no CO\\
14& J195832.30+313304.2 & 68.238, 1.129 & 13.383& 12.727 & 8.000 &5.588 & Class II, Cloud 4\\
15& J195854.39+311831.1 &68.073, 0.935& 10.025&  9.456 &  6.587 & 4.210 & Class II, no CO\\
\hline
\multicolumn{8}{c}{{\bf Herschel sources}} \\
$\#$ &  Designation &  {(\it l, b})& F$_{70}$ [Jy]  &   F$_{160}$ [Jy]   &  F$_{250}$ [Jy]  &  F$_{350}$ [Jy] & F$_{500}$ [Jy] \\
\hline
16 & 238687 &  67.973, 1.022 & 0.329 &0.761 & 1.611 & --& --\\
17& 238749 &68.008, 1.049 & 0.785&  1.689 &5.746& 4.558& 2.004\\
18& 238786 & 68.026, 1.062 & 1.3 & 2.864 & 6.511 &12.94 &2.957\\
19& 238789 & 68.028, 0.983 & 1.276 &5.812 &5.762 &3.265 & 1.663\\
20& 238797 & 68.033, 1.054 & 11.424 & 16.491 &12.655& 2.347& 2.293\\
21& 238835& 68.057, 1.076 & 0.455& 3.849& 6.204&  --& --\\
22& 238863& 68.069, 1.062 &15.427&20.963 &13.061 &10.912 &6.197\\
23& 238880& 68.078, 1.074 &74.122& 49.372 &36.245& 18.919 &9.604\\
24& 238886& 68.08,  0.959&0.804 & 3.267 &2.641& 1.705& 2.219\\
25& 238887& 68.08, 0.966&0.891& 2.022& 4.1 & --& --\\
26& 238888& 68.080, 1.06 &1.053 &3.049& 1.564& --& --\\
27& 238953 &68.124, 0.939 &1.478& 4.173 & 3.185 & 5.83& 2.375\\
28& 238963 &68.130, 0.921& 0.551 &1.354&2.001 & --& --\\
29& 238979& 68.138, 0.933&2.115 &3.978&1.344& --& --\\
30& 238993 &68.143, 0.897& 8.92& 14.102&13.491 &0.501 &--\\
31&  239001& 68.145, 0.905 &1.717 &10.201& 6.913& 12.91& 6.887\\
32& 239003& 68.146, 0.922 &5.407&27.825&-- &--& --\\
33& 239007 &68.148, 0.897&1.284&3.301&-- &--& --\\
34& 239008& 68.148, 0.915&3.876 &3.983&16.76&--& --\\
35&239012&  68.150, 0.928&16.703&70.192 &35.56&--& --\\
36&  239013& 68.150, 0.922 &8.439&46.275& 35.76& 21.896 &11.192\\
37&  239094 &68.195, 1.125& 2.781 &13.222&9.275 &4.634 &2.413\\
38& 239192 &68.24, 0.959&37.411&18.088&15.479& 5.295& 2.463\\
39& 239252& 68.266, 1.116&1.249& 4.231&2.361&--& --\\
40& 239271 &68.276, 1.185&0.955 &1.444&1.758& 0.639&--\\
41& 239279& 68.279, 0.948&0.739 &6.536& 12.99& 8.077 &4.812\\
42& 239302 &68.294, 1.065&0.73 &1.843&6.333& 1.531&1.218\\
43& 239366 &68.323,  1.076&1.567 &13.42& 7.726 &11.75 &4.159\\
44& 239370 &68.327, 1.075&5.431 &11.735&16.903 &--& --\\
45&  239371 & 68.328, 1.02&4.212& 0.872&0.808&--& --\\
\hline
\end{tabular}
\end{table*}

The location of all the cYSOs found  are indicated in Fig. \ref{figysos} and their infrared  properties listed in Table \ref{ysos}.
From Fig. \ref{figysos} it is evident that star formation is taking place in the local environment of WR\,130, and that the cYSOs are found preferentially onto the molecular clouds labeled as Cloud 1, 2, 3, and 4, in Fig. \ref{set-co}.
It is clear that the spatial distribution of the cYSOs follows the distribution of the high column density matter.

As can be noticed, the IRAS source IRAS\,19571+3113 (\# 2 in  Table \ref{ysos}), two MSX sources (\#4 and \#7), and 9 Herschel sources (\#28 to \#36) are located onto the \hii\, region G68.14+0.92, which, as pointed out by \citet{cic01}, is not related to WR\,130. Moreover, \citet{bro96} in their survey of the CS(2-1) emission toward IRAS point sources in the galactic plane, observed, in direction to IRAS\,19571+3113, CS(2-1) emission at --62.6 \kms, suggesting that a high density molecular clump is associated with  G68.14+0.92 and may harbor new stars. Given that this region is far away from the area under study, these  12 infrared sources will not be considered further.

As regards to the MSX source candidate to be a CHII G068.2383+00.9596 (\# 5 in Table \ref{ysos}),  it coincides with the Herschel source \#38 and, from Fig. \ref{figysos}, it can be seen  that, in projection,  their location is very close to  WR\,130. This source is observed both at radio and infrared wavelengths and it is included in the Red MSX Source (RMS) survey \citep{ur208}. According to these authors, G068.2383+00.9596 is associated with  the infrared source IRAS\,19572+3119, several 2MASS point sources, the WISE source 5087 (with estimated flux values only at 12 and 22 $\mu$m), and the radio source G068.2389+00.9592 \citep{urq09}, for which  a flux intensity of 6.1 mJy is given at 5 GHz.
On the other hand, \citet{urq11}  detect neither ammonia nor water maser emission in  the direction to this infrared source, and cannot thus suggest any kinematical distance to it. An inspection of the CO data cube shows, however, that CO emission in this direction is only  detected at about --62 \kms, as shown in Fig. \ref{msx}, suggesting a further distance for the CHII candidate, of about 12 kpc, similar to the distance of the \hii\, region G68.14+0.92, indicating that  we have located an active star formation region that is located  far beyond the solar circle. In this case, the infrared sources \#5 and \#38 can not be associated with WR\,130.  
On the other hand, an inspection of the entire CO data cube reveals that sources \#13 and \#15 (see Table \ref{ysos}) are not immersed in any molecular cloud,  indicating that, if they  are YSOs,  either  they have already moved away from the place where they were formed or that they dissipated it.

\begin{figure}
\centering
\includegraphics[width=9cm]{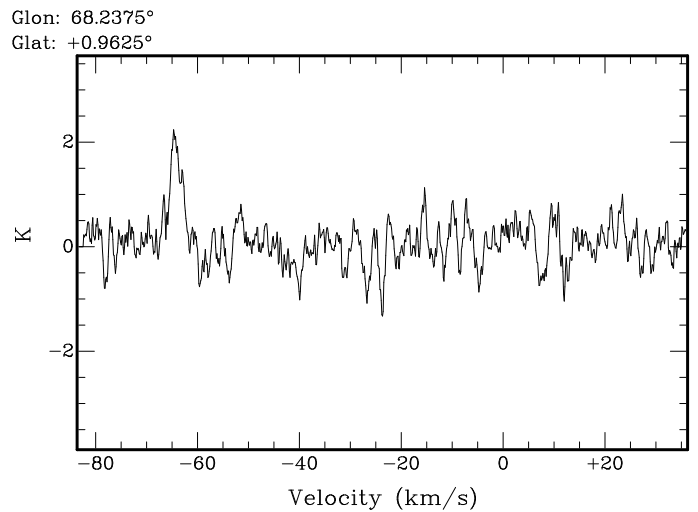}
\caption{CO emission in direction to the MSX source  G068.2383+00.9596.}
\label{msx}
\end{figure}

From the remaining cYSOS appearing in Table \ref{ysos}, it is important to mention that CO emission in their direction is only observed at the radial velocity interval between --9 and --15 \kms, coincident with the radial velocity of the molecular structure, suggesting an association between them and the molecular clouds. The  exceptions  are Source \#8, for which CO emission is also detected at  --60 \kms\,  and Sources \#22 and \#26, in which direction there is CO emission also at $\sim -40$ \kms.

Among the cYSOs, two groups of  sources 
 are of particular interest.
As can be seen from Fig. \ref{figysos},  four sources,  
\#1-23 (since sources \#1 and \#23 spatially coincide), \#21, \#22, and \#26
are seen projected
on the periphery of the CO cloud named as Cloud 1  in Fig. \ref{set-co},  while three sources,
 \#3-12-20, \#17 and \#18, are locatated onto Cloud 2. In both cases the cYSOs are located  on the side nearly facing the star
WR\,130.  This suggests that they could be  bright-rimmed clouds
\citep[e.g.,][]{mor09} being excited by the ionizing radiation
from WR\,130, in which case one could expect to find
a radio continuum counterpart \citep[e.g.,][]{sch85}. An inspection of Fig. \ref{br} shows indeed the presence of a weak radio continuum source 
possibly barely resolved (it is hard to clearly separate it from its 
surrounding emission) with the CGPS beam, 
thus suggesting a size of approximately 1\arcmin or less.
A peak 1420 MHz flux density of $\sim 10 \pm 1$ mJy is obtained with an estimated
total flux density of $18 \pm 3$ mJy
in the border of Cloud 1 and a peak flux value of about 2 mJy is estimated for the emission bordering Cloud 2.

\begin{figure}
\centering
\includegraphics[width=9cm]{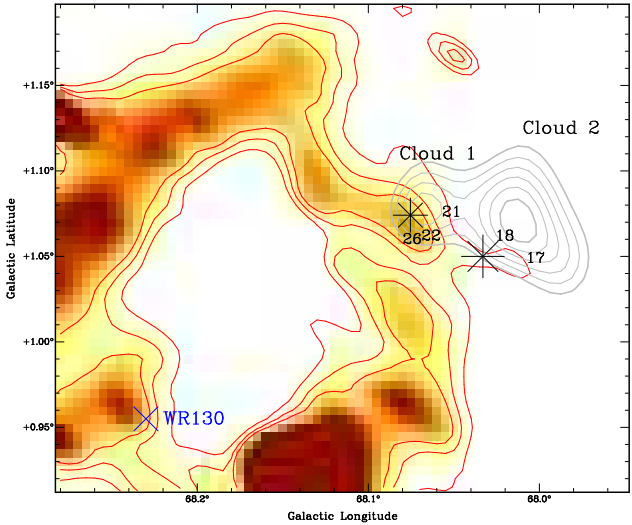}
\caption{ 1420 MHz emission distribution in the area of Sources \#1-23 and \#3-12-20 (both indicated by black asterisks). Red contours are at 9.2, 9.6 and 10 K. Black contours correspond to the averaged CO emission between --2.35 and 4.25  \kms, from 1 to 2 K with a step of 0.2 K. The location of WR\,130 is indicated by the blue cross.  Herschel cYSOs are indicating by their source numbers (see Table \ref{ysos}).}
\label{br}
\end{figure}

\begin{table*}
\caption{Estimated parameters for the boundary ionized layers related to clouds 1 and 2 .} 
\label{rdi}
\centering  
\begin{tabular}{c  c c c c c  c}
\hline \hline
 Cloud &  $n_e$ (cm$^{-3}$) & $M_i$ ($M_{\odot}$) & $N_{u}$ ($10^{46}$\, s$^{-1}$) &$N^{i}_{\rm WR}$ ($10^{46}$\, s$^{-1}$) & $P_i/k$ ($10^6$ cm$^{-3}$\,K) & $P_m/k$ ($10^6$ cm$^{-3}$\,K)\\
\hline
1 &  29&2.2& 2.0& 3.0 & 1.9  & 2.0\\
2  &  12 & 1 & 0.4 & 3.5 & 0.6 & 6\\
\hline
\end{tabular}
\end{table*}

Assuming the radio continuum emission to be free-free emission, we
can use the
models of \citet{mez67} to infer the properties
of the ionized gas, following the procedure outlined by
\citet{pin98}.  Since we do not know a priori the actual
distribution of the gas, we take the simplest possible model for the
source, namely that of a sphere with constant density.

Then the electron density $n_e$,
mass of ionized gas $M$ and total number $N_u$ of UV ionizing 
photons per second \citep[e.g.][]{cha76} are given  by:

$$n_e  =  6.35 \times 10^2 \,u_1\,a^{-1/2}\,T_4^{0.175}\,\nu_{\rm GHz}
           ^{0.05}\,S_{\nu}^{0.5}\,\theta_{\rm Ga}^{-1.5}\,D_{\rm kpc}
           ^{-0.5} \rm\, cm^{-3},$$
$$M/M_{\odot}  =  0.386\,u_2\,a^{-1/2}\,T_4^{0.175}\,\nu_{\rm GHz}
           ^{0.05}\,S_{\nu}^{0.5}\,\theta_{\rm Ga}^{1.5}\,D_{\rm kpc}
           ^{2.5},$$
$$N_u  =  0.76 \times 10^{47}\,\,T_4^{-0.35}\,\nu_{\rm GHz}
           ^{0.1}\,S_{\nu}\,\,D_{\rm kpc}{^{2}} \rm \, s^{-1},$$

\noindent where $S_{\nu}$ is the measured flux density in Jy, $\theta_{Ga}$ 
the measured half-power width in minutes of arc, $T_4$ the electron temperature 
in units of $10^4$ K,  $a(\nu,T_e)$ the Gaunt factor (of order unity),
$D_{\rm kpc}$ the distance in kpc and $\nu_{\rm GHz}$ the frequency in GHz.
The parameters $u_1$ and $u_2$ are of order unity for a spherical
source model.
As the observing frequency appears with a very low exponent, all frequency
factors are also of order unity.  Assuming a distance of 5 kpc \citep{cic01}, a 
temperature of $10^4$ K and using $\theta_{\rm Ga} \sim 1$ 
(valid for a source unresolved with a beam of the order of 1\arcmin),
the above three equations thus simplify to

$$n_e  \sim  290 \, S_{\nu}^{0.5} \rm\, cm^{-3},$$
$$M/M_{\odot}  \sim  22 \,S_{\nu}^{0.5}, $$ 
$$N_u  \sim  1.9 \times 10^{48} \,S_{\nu}\,\rm \, s^{-1}.$$

It should be emphasized at this point that the above estimate for $n_e$ is likely an underestimate since the source geometry is that of a thin layer of ionized gas on the star-facing side of the cloud.  Since the electron number density is of the form $n_e \propto \theta_{\rm Ga}^{-1.5} \propto {\rm (volume)}^{-1/2}$, we expect $n_e$ to scale roughly as $\eta^{-1/2}$ if we assume the thickness of the layer to be given by  $\eta$ times the cloud radius.

Using the measured peak flux density of 10 mJy and 2 mJy for the border of cloud 1 and 2 respectively, the above numbers reduce to the values given in Table \ref{rdi}.
As both sources are essentially unresolved at 1\arcmin\, 
the estimated densities and ionized masses are a lower and upper limit, respectively. 
Their corresponding maximum size is on the order of 1.5 pc for a distance of 5 kpc.

Assuming that WR\,130 is the star providing the ionizing radiation, which
\citet{van01} gives as $N_{\rm WR} = 5.0 \times 10^{48}  \rm \, s^{-1}$,
we can calculate the number of ionizing photons $N^i_{\rm WR}$ intercepted
by the radio continuum features bordering cloud 1 and 2. For this estimate we considered that the border of cloud 1
 is located some 12\arcmin\, or 17.4 pc from WR~130 (we do not make any
correction to the distance or cloud size due to projection effects as these
cancel out), and the border of cloud 2 is a little bit further from WR\,130, at about 12\farcm5 or 18 pc.
We find that a fraction $f1 \sim 0.006$ and $f2 \sim 0.007$ of the photons from
WR\,130 would be intercepted for each cloud, so that $N^{i1}_{\rm WR} \sim 3.0 \times 10^{46}\rm\, s^{-1}$ and $N^{i3}_{\rm WR} \sim 3.5 \times 10^{46}\rm\, s^{-1}$ .
It would seem, at least at first sight, that WR\,130 alone is sufficient to power 
the observed radio continuum emission.

We thus have two molecular clouds, labeled Cloud 1 and 2,  presenting a bright rim and each harboring  cYSOs. 
This together with the fact that the observed cometary shapes of the bright rims are  quite apparent with their tips pointed towards the location of WR\,130 (see Fig. \ref{figysos}) suggest the possibility  that the cYSOs were formed under the radiative driven implosion (RDI) mechanism \citep{ber89}. This mechanism takes place when  a photoionization shock is driven into the molecular gas. To ascertain whether this is the case, it is necessary to compare the pressure of the ionized layer bordering the molecular cloud with the molecular pressure. If the cloud is under-pressured with respect to the ionized layer or the pressures are similar, the ionization front will be able to propagate into the cloud and then modify its evolution \citep{lef94}.

The pressure in the ionized boundary layer can be evaluated from the electron density as $P_i = 2\, \rho_i\, c_i^2$, where $\rho_i$ is the density in the boundary layer,  $\rho_i = n_e \, m_H$ and $c_i$ is the sound speed in the ionized gas ($c_i \sim 11.4$ \kms). 
As mentioned above, this pressure should be compared with the pressure of the molecular cloud, which can be estimated, assuming that the thermal component can be neglected, from the turbulent velocity dispersion, $\sigma^2$, and the molecular density, $\rho_m$, as $P_m = \sigma^2 \, \rho_m$, where  $\sigma^2$ may be written as $\sigma^2 = \Delta\, v^2/$(8\, ln 2), $\Delta\,v$ being the observed velocity line width of the molecular cloud gas.
The results are given in Table \ref{rdi}.
Before comparing the obtained values it is important to mention that the errors involved in these estimates are large. As mentioned before, the electron densities obtained for the ionized gas are lower limits, and thus the estimates of $P_i$ are also underestimates.
On the other hand, since the molecular pressure depends on the geometry adopted to estimate the molecular density, the CO-to-H$_2$ conversion factor $X$, and the distance adopted for the structure, its value has at least a 60\% error.
Bearing this in mind, from Table \ref{rdi} we can see that the pressures $P_i$ and  $P_m$ obtained for cloud 1 are similar while in the case of cloud 2 the molecular pressure is higher than the pressure of the ionized gas, suggesting that the shock could not yet affect the molecular gas. However, given the large errors involved in the estimates, with the data at hand we cannot completely rule out the possibility that the shock front has already penetrated into  cloud 2.

The second group of sources of interest in Fig. \ref{figysos} is located onto Cloud 4 (see Fig. \ref{set-co}). This group consists of five WISE sources  (two of them coincide with Herschel sources), two classified as Class I YSOs and three as more evolved, Class II, objects (see Table \ref{ysos}). An inspection of Fig. \ref{wh} clearly shows, more strikingly at 250 $\mu$m, two jet-like features spatially coincident with the border of molecular Cloud 4. Moreover, two of the cYSO, identified as sources \#9, \#39  and \#14 in Table \ref{ysos}, are located onto the tip of these structures. A closeup of this region is shown in Fig. \ref{cloud4}. The cometary shape of the infrared structures suggests that the RDI mechanism could be at work in this molecular cloud, but the low angular resolution of both the molecular and radio continuum data prevents carrying out the analysis in this region.

 Among the rest of the cYSOs presented in Table \ref{ysos}, it can be seen that four sources (\#6-45, \#42, \#43, and \#44) are located onto Cloud 3 (see Fig. \ref{set-co}), while the rest are distributed onto the general CO shell structure.

\subsection{Physical Properties of cYSOs}

To characterize the cYSOs identified located in projection onto the molecular clouds, we constructed their SED  using the grid of models and fitting tools of \citet{rob06,rob07}.
The SED fitting tool fits the data allowing the distance and external foreground extinction as free parameters. We gave a distance range of 4.5 - 5.5 kpc. The
visual absorption range for each source was set  making use of the \hi\, and CO data cubes to estimate  their column densities
at the distance of the structure, using the relation N(HI) + 2 N(H$_2$) $\sim 1.9 \times 10^{21} A_V$ \citep{boh78}, where N(HI) is the \hi\, column density and is given by 
N(HI)(cm$^{-2}$) = $1.824 \times 10^{18}\, T ({\rm K})\, \Delta v$ (\kms), and N(H$_2$) is the molecular column density, given by N(H$_2$)(cm$^{-2}) = 2.3 \times 10^{20}\,$ T (K)\, $\Delta v$ (\kms), where, in both cases,  $T$ is the brightness temperature and $  \Delta v$ is the velocity interval between the structure and the observer. The adopted range for each source is given in the last column of Table \ref{seds}.

For each cYSO besides using the photometry given in Table \ref{ysos} we search their counterpart in other infrared ranges, to use in the fitting as many points as possible to better constrain the estimated parameters.
Thus, for sources \#3-12-20, \#11, \#14 and \#15, JHK photometry was obtained from the 2MASS catalogue.
For source \#1-23 the flux given in the MSX catalog at Band A was also considered.
For source \#3-12-20 out of the four MSX bands observed only Band A was considered due to the bad quality of the other three, and the IRAS fluxes obtained from the IRAS catalogue were used as upper limits.

For each source, based on  all  the fitted models satisfying $\chi^2 - \chi^2_{\rm best} < 2\, N$, where $\chi^2_{\rm best}$ is the goodness-of-fit parameter for the  best-fit model  and  $N$ is the number of input observational data points, we estimated  some parameters such as the stellar mass, stellar temperature, stellar age,  envelope accretion rate, envelope mass and total source luminosity. 
To estimate the parameters,  we carefully inspect every distribution  considering all the models that satisfy the criteria mentioned above. The results are listed in Table \ref{seds}. 
We note that, with the data at hand, even when  the data points can be fitted with  great goodness, just a few parameters are well constrained and in some cases only  lower and/or upper limit values can be given.

\begin{figure}
\centering
\includegraphics[width=9cm]{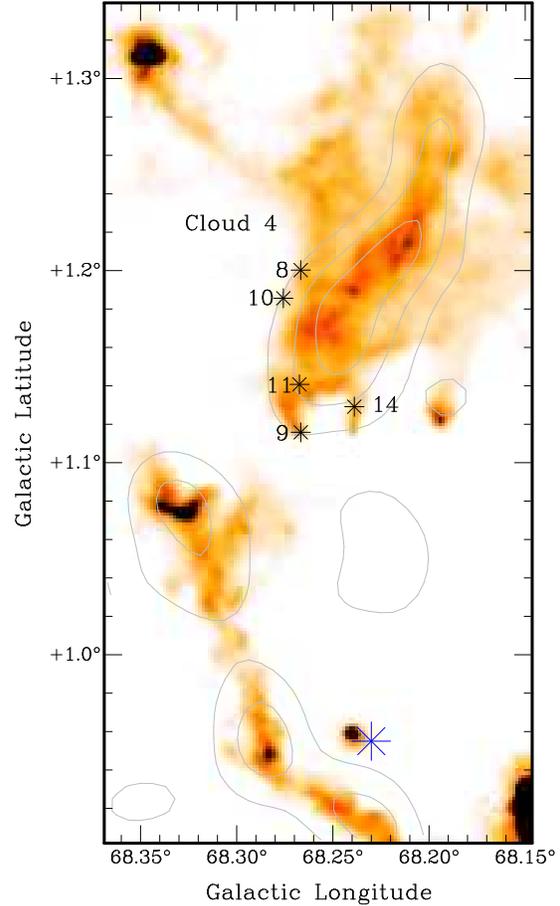}
\caption{ 250 $\mu$m emission distribution in the area of the molecular cloud 4. The five WISE sources located onto this cloud are indicated with asterisks and with their corresponding numbers (see Table \ref{ysos}). Gray contours are from 0.4 to 1.6 K with a step of 0.4 K and correspond to the averaged CO emission between --7.18 and --0.7  \kms. The location of WR\,130 is indicated by the blue asterisk. }
\label{cloud4}
\end{figure}

\begin{table*}
\caption{ SED parameters obtained for the cYSOs. \label{seds}}
\centering
\begin{tabular}{c c c c c c c c }
\hline
\hline
$\#$ &    Stellar Mass &  Stellar Temperature &   Stellar Age  &     Envelope mass &Envelope Accretion Rate & Total Luminosity & A$_V$      \\
& $M_{\odot}$ &  K &  $10^3$  yr & $M_{\odot}$& $M_{\odot}\, \rm yr^{-1}$ & $10^3$ $L_{\odot}$ & mag\\
\hline
1-23 & 8 & 4130& 1.3 &476& 5.6 $\times 10^{-4}$ & 2.5 & 4--10\\
3-12-20 & 4  & 4143& 1.7 &20 & 0.002 & 0.3 & 4 --8\\
6-45 &6.8 & 4435& 21 & 0.4& 3.7 $\times 10^{-5}$&0.9 & 2 -- 5\\
8 & 3 -- 6 & 11000 -- 18700 & $>1200$ & $<7 \times 10^{-7}$& 0 & 0.05--0.97 &2 --4\\
9-39 & 1.2 & $ 3967 \pm 15 $ & 1.1--1.3 & 69--72 &($0.4 \pm 0.1$)$ \times 10^{-4}$& 0.04--0.05 &2 -- 6\\ 
10-40& 4.6 &  11270 &836 &10.6 & 1.5 $\times 10^{-7}$& 0.4 & 3 --7\\
11 & $5.3 \pm 0.4$ & 4688 $\pm 118$ & 95--185 &8 -- 13& (0.3 -- 1.0) $\times 10^{-4}$&0.1--0.2 & 2 -- 6\\
13 & 4.2 $\pm 0.3$ & 14855 $\pm 600$& 2200 -- 4100&  $< 9 \times 10^{-8}$&0 &0.26 $\pm 0.06$ & 2 -- 6 \\
14 &  1.9 & 4320 &112 & 12 &9 $\times 10^{-5}$& 36 & 2 -- 7\\
15&3.5 &11820 &1954 & $10^{-8}$& 0  & 0.2& 2 --4 \\
16 & 1.3 $\pm$ 0.3&  4092 $\pm$ 94& 7-- 14& 97 $\pm$ 17&  (50 $\pm$$ 7) \times 10^{-5}$& 0.03 $\pm$ 0.01 & 5 --8 \\
17& 3.4& 4390&39 &251 & 10$^{-3}$& 0.09 & 6 --9\\
18& 5.5& 4433&28 &760 & 2 $\times $10$^{-3}$& 0.2 & 7 --10\\
19& 5.5& 4433& 28& 760& 2 $\times $10$^{-3}$& 0.2 & 4 --7   \\
21& 2.6& 4200&7.6 & 274& 10$^{-3}$&  0.1 & 6 --9\\
22& 2.3 -- 2.5& 4215 $\pm$ 22&8 --15  & 223 -- 274&  10$^{-3}$& 0-07 -- 0.10 &6 --9\\
24& 2.3 -- 3.4& 4200 -- 4393& 6.4 -- 38.7&223 -- 345 & 10$^{-3}$& 71 -- 112& 4 --7 \\
25& 1.6 -- 3.4&  4114 -- 4393& 6 -- 39& 165 -- 251&  10$^{-3}$ &0.05 -- 0-09& 4 --7\\
26& 2.5 $\pm$ 0.8 &4254 $\pm$ 125 & 23 $\pm$ 9 &14 -- 262 & (7 $\pm$ 2) $\times $10$^{-4}$&0.08 $\pm$ 0.02 & 4 -- 8 \\
27& 5.5 & 4433& 28&760 & 2 $\times $10$^{-3}$&  0.2 & 3 -- 5\\
37& 4.9&4380 & 6 -- 20&  540 -- 840& 2 $\times $10$^{-3}$ & 0.25 -- 0.30 & 4 -- 6\\
41& 4.6& 4208& 4.7& 1241 & 2 $\times $10$^{-3}$& 0.3 & 5 -- 8 \\
42& 2.3 -- 3.4& 4200 -- 4393& 7.6 -- 38.8& 223 -- 274&10$^{-3}$ & 0.07 -- 0.10 & 2 -- 5\\
43&4.6 & 4208& 4.7& 1241&2 $\times $10$^{-3}$ &0.3 & 5 -- 8 \\
44& 4.6 -- 8.1&4423 $\pm$ 189 &$<$ 30 & 1142 $\pm$ 228& 2 $\times $10$^{-3}$& 0.5 $\pm$ 0.2 & 5 -- 8\\
\hline
\end{tabular}
\end{table*}
 
As can be seen in Table \ref{seds} all the sources but  five are in a very early stage of their formation, with ages lower than $10^5$ yr,  still having high envelope masses, of the order of hundreds, and envelope accretion rates in the order of $10^{-5} - 10^{-3}$ $M_{\odot}\, \rm yr^{-1}$.
Similarly, most of the cYSOs have low/intermediate masses, lower than 7  $M_{\odot}$, although these values will probably increase considerably since they are still accreting lot of mass from their envelopes. 

 As mentioned before, Fig. \ref{figysos} shows that sources \#1-23, \#21, \#22, and \#26 are located  within the molecular cloud 1, while sources \#3-12-20, \#17 and \#18 are within cloud 2.
An inspection of the ages estimated for these cYSOs shows that in cloud 1 the age gradient is as expected for sources \#1-23, \#22 and \#26, if the RDI mechanism took place in this cloud, in the sense that the oldest source (\#26, among these three) is located closest (in projection) to the WR star, then is source \#22 and then \#1-23. The problem is with source \#21, which, since it is located further from WR\,130  should be younger. 
A similar situation occurs for the three sources located within cloud 2, where the ages of sources \#17 and \#18 are larger than the one of \#3-12-20.
Looking at the SEDs fitting  we found that for these three sources particulary the  fitting is not  satisfactory. More data at other wavelengths would be useful to better  estimate the  age of these sources.

Three of the sources that have higher ages are sources \#8, \#10-40 and  \#11, all lying in projection onto Cloud 4. In this case, if the RDI mechanism is  responsible for the star formation in the cloud, given the location of these three cYSOs with respect to WR\,130, they would have the lowest ages. Since this is not what we obtained, we suggest that either  another mechanism is acting in cloud 4 or that both sources are not related to this cloud. As mentioned before, CO emission is also detected at --60 \kms\, in  direction to source \#8. 
Also sources \#13 and \#15 are among the oldest cYSOs, which is consistent with the fact that, as mentioned before, they are not inmersed in any molecular gas.

The simultaneous presence of envelope-dominated YSOs (typical age around 10$^4$ years) and a WR star in the same region suggests that different episodes of star formation have occurred in the region.
Since WR\,130 is an evolved star, we can speculate that the formation of young sources in the region might have been triggered by the expanding bubble.

\section{Summary}

 The previous observations by \citet{cic01} and those reported here allow us to
describe a consistent picture of the recent evolution of WR\,130 and of its 
effect on its immediate environment on large as well as small scales.
The intitial results of \citet{cic01} showed that the continuum radio
emission and \hi\ kinematical analysis were in total agreement with the
generally accepted picture of stellar wind evolution \citep[e.g.][]{cas75, wea77}.
In this paper we have gone one step further and used recent high-quality
IR and CO data to study for the first time the molecular and dust properties of
the ISM local to this Wolf-Rayet star.  

The timescales involved are consistent with the wind of the O star phase
having formed an initial cavity in the surrounding ISM.  The wind of the star
now in its WR phase seems to be interacting directly with the cavity wall as there is
no sign of massive ejection having taken place in the recent history of
the star.  If any such ejection took place, it is likely to have been a
minor event.  In all probability WR\,130 is now in a pre-LBV phase.

On large scales, an inspection of the CO data cube reveals the presence of several molecular 
clouds which have a good morphological agreement with the ionized gas. The CO
structures are observed within the velocity range  between --11 and 3 \kms\,  compatible 
with the velocity of the \hi\  structure and ionized gas as obtained  by \citet{cic01}. 
Among the  observed molecular clouds, four  
have been identified and characterized, over which several cYSOs are located.

At infrared wavelengths, the large-scale distribution of the  WISE emission at 12 $\mu$m clearly 
shows the presence of a PDR bordering  the \hii\, region. 
In Herschel  bands a shell like structure is noticeable, showing a good morphological coincidence  
with the CO structures, suggesting that the dust coexists
with  the molecular gas. This is consistent with dust temperatures of about 24 K 
obtained for the shell structure.

On smaller scales, we see 
clear evidence of gas and dust being in the process of forming new stars, together
with an accompanying erosion of the associated cloud, as a result of the combined 
action of the stellar UV radiation and wind of WR\,130.
This appears most clearly on
the high-resolution WISE images which
show morphological features (e.g., oriented pillars or thin filaments) 
generally seen around young massive stars
or stellar clusters and which are indicative of stellar formation.  

The presence of compact sources at the tip of some of these
structures identifies these as probable YSOs.  Two groups
of such objects have been identified.  In the first case, two of the compact IR sources have an
associated weak radio source located on the star-facing side of their 
associated CO cloud. Using the available information, we have shown that
these two cases are consistent, one formally and the second one marginally,
with a formation via the RDI mechanism.
The other sources, which have a more elongated or pillar-like
appearance, do not have associated radio continuum counterparts and
a similar analysis cannot be carried out. A different mechanism
involving the curvature of a dense expanding shell 
leading to the formation of stable pillar-like structures
has been proposed  \citep[][]{tre12a, tre12b, tre13}
and could very well be
a promising avenue to explain the observed features.  However
such an analysis is clearly out of the scope of this paper.

Based on different colour criteria, an analysis of the
YSO candidates in the vicinity of WR\,130 has allowed us to estimate 
some of their physical parameters and  revealed that most are in a very early phase 
of their formation, still accreting mass, and that they are
mainly low mass stars.  An estimate of their age shows that,
whereas the age gradient is as expected in some cases, there are nevertheless 
cases where the situation is uncertain. It is not clear whether this effect 
is real or the result of a poor fitting of the SEDs.

In summary, the morphological correlation between ionized and neutral gas 
as well as the ring-like dust emission confirm the high
degree of interaction between the wind and UV photons from the WR star and the 
surrounding molecular clouds.  
In addition, the presence of numerous cYSOs  within the molecular gas suggests 
that an active star formation process is  taking place in the environs of this star.

\section*{Acknowledgments}
We thank Sergio Molinari and the HiGAL Consortium for making available
to us the WR130 Herschel data. It is also a pleasure to thank Davide Elia and 
Eugenio Schisano (HiGAL Consortium members), for their help on the
source extraction around the WR130 region, using the Herschel data.
We thank Tony Marston, the referee, for his careful reading
of the manuscript and constructive comments that helped us
to improve its substance \& presentation.
The CGPS is a Canadian Project with international partners
and is supported by grants from NSERC. Data from the CGPS are publicly available through the facilities of the Canadian Astronomy Data Centre (http://cadc.hia.nrc.ca) operated by the Herzberg Institute of Astrophysics, NRC.
We acknowledge  Dr. Christ Brunt who kindly made available the FCRAO CO data to us.
This project was partially financed by the Consejo Nacional de Investigaciones
CientÃ­ficas y TÃ©cnicas (CONICET) of Argentina under project PIP 01299,
Agencia PICT 00902,  and UNLP G091.
L.A. Suad is a postdoctoral fellow of CONICET, Argentina. S. Cichowolski and E.M. Arnal are members of the
Carrera del Investigador CientÃ­fico
of CONICET, Argentina.

\bibliographystyle{mn2e}  
\bibliography{bib-todo}

\begin{thebibliography}{73}
\expandafter\ifx\csname natexlab\endcsname\relax\def\natexlab#1{#1}\fi

\bibitem[{{Anderson} {et~al}\mbox{.}(2012){Anderson}, {Zavagno}, {Deharveng},
  {Abergel}, {Motte}, {Andr{\'e}}, {Bernard}, {Bontemps}, {Hennemann}, {Hill},
  {Rod{\'o}n}, {Roussel}, \& {Russeil}}]{and12}
{Anderson} L.~D. {et~al.}, 2012, \aap, 542, A10

\bibitem[{{Beichman} {et~al}\mbox{.}(1988){Beichman}, {Neugebauer}, {Habing},
  {Clegg}, \& {Chester}}]{bei88}
{Beichman} C.~A., {Neugebauer} G., {Habing} H.~J., {Clegg} P.~E., {Chester}
  T.~J., eds., 1988, {Infrared astronomical satellite (IRAS) catalogs and
  atlases. Volume 1: Explanatory supplement}, Vol.~1

\bibitem[{{Benjamin} {et~al}\mbox{.}(2003){Benjamin}, {Churchwell}, {Babler},
  {Bania}, {Clemens}, {Cohen}, {Dickey}, {Indebetouw}, {Jackson}, {Kobulnicky},
  {Lazarian}, {Marston}, {Mathis}, {Meade}, {Seager}, {Stolovy}, {Watson},
  {Whitney}, {Wolff}, \& {Wolfire}}]{ben03}
{Benjamin} R.~A. {et~al.}, 2003, \pasp, 115, 953

\bibitem[{{Bertoldi}(1989)}]{ber89}
{Bertoldi} F., 1989, \apj, 346, 735

\bibitem[{{Bohlin}, {Savage} \& {Drake}(1978){Bohlin}, {Savage}, \&
  {Drake}}]{boh78}
{Bohlin} R.~C., {Savage} B.~D., {Drake} J.~F., 1978, \apj, 224, 132

\bibitem[{{Boulanger} {et~al}\mbox{.}(1996){Boulanger}, {Abergel}, {Bernard},
  {Burton}, {Desert}, {Hartmann}, {Lagache}, \& {Puget}}]{bou96}
{Boulanger} F., {Abergel} A., {Bernard} J.-P., {Burton} W.~B., {Desert} F.-X.,
  {Hartmann} D., {Lagache} G., {Puget} J.-L., 1996, \aap, 312, 256

\bibitem[{{Bronfman}, {Nyman} \& {May}(1996){Bronfman}, {Nyman}, \&
  {May}}]{bro96}
{Bronfman} L., {Nyman} L.-A., {May} J., 1996, \aaps, 115, 81

\bibitem[{{Cappa} {et~al}\mbox{.}(2010){Cappa}, {Vasquez}, {Pineault}, \&
  {Cichowolski}}]{cap10}
{Cappa} C.~E., {Vasquez} J., {Pineault} S., {Cichowolski} S., 2010, \mnras,
  403, 387

\bibitem[{{Carey} {et~al}\mbox{.}(2009){Carey}, {Noriega-Crespo}, {Mizuno},
  {Shenoy}, {Paladini}, {Kraemer}, {Price}, {Flagey}, {Ryan}, {Ingalls},
  {Kuchar}, {Pinheiro Gon{\c c}alves}, {Indebetouw}, {Billot}, {Marleau},
  {Padgett}, {Rebull}, {Bressert}, {Ali}, {Molinari}, {Martin}, {Berriman},
  {Boulanger}, {Latter}, {Miville-Deschenes}, {Shipman}, \& {Testi}}]{car09}
{Carey} S.~J. {et~al.}, 2009, \pasp, 121, 76

\bibitem[{{Castor}, {McCray} \& {Weaver}(1975){Castor}, {McCray}, \&
  {Weaver}}]{cas75}
{Castor} J., {McCray} R., {Weaver} R., 1975, \apjl, 200, L107

\bibitem[{{Chaisson}(1976)}]{cha76}
{Chaisson} E.~J., 1976, in Frontiers of Astrophysics, {Avrett} E.~H., ed., pp.
  259--351

\bibitem[{{Cichowolski} {et~al}\mbox{.}(2001){Cichowolski}, {Pineault},
  {Arnal}, {Testori}, {Goss}, \& {Cappa}}]{cic01}
{Cichowolski} S., {Pineault} S., {Arnal} E.~M., {Testori} J.~C., {Goss} W.~M.,
  {Cappa} C.~E., 2001, \aj, 122, 1938

\bibitem[{{Compi{\`e}gne} {et~al}\mbox{.}(2011){Compi{\`e}gne}, {Verstraete},
  {Jones}, {Bernard}, {Boulanger}, {Flagey}, {Le Bourlot}, {Paradis}, \&
  {Ysard}}]{com11}
{Compi{\`e}gne} M. {et~al.}, 2011, \aap, 525, A103

\bibitem[{{Conti}, {Leep} \& {Perry}(1983){Conti}, {Leep}, \& {Perry}}]{con83}
{Conti} P.~S., {Leep} M.~E., {Perry} D.~N., 1983, \apj, 268, 228

\bibitem[{{Crowther}(2007)}]{cro07}
{Crowther} P.~A., 2007, \araa, 45, 177

\bibitem[{{de Jager}, {Nieuwenhuijzen} \& {van der Hucht}(1988){de Jager},
  {Nieuwenhuijzen}, \& {van der Hucht}}]{jag88}
{de Jager} C., {Nieuwenhuijzen} H., {van der Hucht} K.~A., 1988, \aaps, 72, 259

\bibitem[{{Deharveng} {et~al}\mbox{.}(2010){Deharveng}, {Schuller}, {Anderson},
  {Zavagno}, {Wyrowski}, {Menten}, {Bronfman}, {Testi}, {Walmsley}, \&
  {Wienen}}]{deh10}
{Deharveng} L. {et~al.}, 2010, \aap, 523, A6

\bibitem[{{Dyson} \& {Williams}(1997)}]{dys97}
{Dyson} J.~E., {Williams} D.~A., 1997, {The physics of the interstellar
  medium}. The physics of the interstellar medium.~ Edition: 2nd ed.~Publisher:
  Bristol: Institute of Physics Publishing, 1997.~Edited by J.~E.~Dyson and
  D.~A.~Williams.~Series: The graduate series in astronomy.~ISBN: 0750303069

\bibitem[{{Egan} {et~al}\mbox{.}(2003){Egan}, {Price}, {Kraemer}, {Mizuno},
  {Carey}, {Wright}, {Engelke}, {Cohen}, \& {Gugliotti}}]{ega03}
{Egan} M.~P. {et~al.}, 2003, VizieR Online Data Catalog, 5114, 0

\bibitem[{{Elia} {et~al}\mbox{.}(2013){Elia}, {Molinari}, {Fukui}, {Schisano},
  {Olmi}, {Veneziani}, {Hayakawa}, {Pestalozzi}, {Schneider}, {Benedettini},
  {di Giorgio}, {Ikhenaode}, {Mizuno}, {Onishi}, {Pezzuto}, {Piazzo},
  {Polychroni}, {Rygl}, {Yamamoto}, \& {Maruccia}}]{eli13}
{Elia} D. {et~al.}, 2013, \apj, 772, 45

\bibitem[{{Elmegreen}(1998)}]{elm98}
{Elmegreen} B.~G., 1998, in Astronomical Society of the Pacific Conference
  Series, Vol. 148, Origins, {Woodward} C.~E., {Shull} J.~M., {Thronson} Jr.
  H.~A., eds., p. 150

\bibitem[{{Evans} {et~al}\mbox{.}(2009){Evans}, {Dunham}, {J{\o}rgensen},
  {Enoch}, {Mer{\'{\i}}n}, {van Dishoeck}, {Alcal{\'a}}, {Myers},
  {Stapelfeldt}, {Huard}, {Allen}, {Harvey}, {van Kempen}, {Blake}, {Koerner},
  {Mundy}, {Padgett}, \& {Sargent}}]{eva09}
{Evans}, II N.~J. {et~al.}, 2009, \apjs, 181, 321

\bibitem[{{Fowler} \& {Aumann}(1994)}]{fow94}
{Fowler} J.~W., {Aumann} H.~H., 1994, in Science with High Spatial Resolution
  Far-Infrared Data, {Terebey} S., {Mazzarella} J.~M., eds., pp. 1--+

\bibitem[{{Garcia-Segura} \& {Mac Low}(1995)}]{gar95}
{Garcia-Segura} G., {Mac Low} M.-M., 1995, \apj, 455, 145

\bibitem[{{Griffin} {et~al}\mbox{.}(2010){Griffin}, {Abergel}, {Abreu}, {Ade},
  {Andr{\'e}}, {Augueres}, {Babbedge}, {Bae}, {Baillie}, {Baluteau}, {Barlow},
  {Bendo}, {Benielli}, {Bock}, {Bonhomme}, {Brisbin}, {Brockley-Blatt},
  {Caldwell}, {Cara}, {Castro-Rodriguez}, {Cerulli}, {Chanial}, {Chen},
  {Clark}, {Clements}, {Clerc}, {Coker}, {Communal}, {Conversi}, {Cox},
  {Crumb}, {Cunningham}, {Daly}, {Davis}, {de Antoni}, {Delderfield}, {Devin},
  {di Giorgio}, {Didschuns}, {Dohlen}, {Donati}, {Dowell}, {Dowell}, {Duband},
  {Dumaye}, {Emery}, {Ferlet}, {Ferrand}, {Fontignie}, {Fox}, {Franceschini},
  {Frerking}, {Fulton}, {Garcia}, {Gastaud}, {Gear}, {Glenn}, {Goizel},
  {Griffin}, {Grundy}, {Guest}, {Guillemet}, {Hargrave}, {Harwit}, {Hastings},
  {Hatziminaoglou}, {Herman}, {Hinde}, {Hristov}, {Huang}, {Imhof}, {Isaak},
  {Israelsson}, {Ivison}, {Jennings}, {Kiernan}, {King}, {Lange}, {Latter},
  {Laurent}, {Laurent}, {Leeks}, {Lellouch}, {Levenson}, {Li}, {Li},
  {Lilienthal}, {Lim}, {Liu}, {Lu}, {Madden}, {Mainetti}, {Marliani}, {McKay},
  {Mercier}, {Molinari}, {Morris}, {Moseley}, {Mulder}, {Mur}, {Naylor},
  {Nguyen}, {O'Halloran}, {Oliver}, {Olofsson}, {Olofsson}, {Orfei}, {Page},
  {Pain}, {Panuzzo}, {Papageorgiou}, {Parks}, {Parr-Burman}, {Pearce},
  {Pearson}, {P{\'e}rez-Fournon}, {Pinsard}, {Pisano}, {Podosek}, {Pohlen},
  {Polehampton}, {Pouliquen}, {Rigopoulou}, {Rizzo}, {Roseboom}, {Roussel},
  {Rowan-Robinson}, {Rownd}, {Saraceno}, {Sauvage}, {Savage}, {Savini},
  {Sawyer}, {Scharmberg}, {Schmitt}, {Schneider}, {Schulz}, {Schwartz},
  {Shafer}, {Shupe}, {Sibthorpe}, {Sidher}, {Smith}, {Smith}, {Smith},
  {Spencer}, {Stobie}, {Sudiwala}, {Sukhatme}, {Surace}, {Stevens}, {Swinyard},
  {Trichas}, {Tourette}, {Triou}, {Tseng}, {Tucker}, {Turner}, {Vaccari},
  {Valtchanov}, {Vigroux}, {Virique}, {Voellmer}, {Walker}, {Ward}, {Waskett},
  {Weilert}, {Wesson}, {White}, {Whitehouse}, {Wilson}, {Winter}, {Woodcraft},
  {Wright}, {Xu}, {Zavagno}, {Zemcov}, {Zhang}, \& {Zonca}}]{gri10}
{Griffin} M.~J. {et~al.}, 2010, \aap, 518, L3

\bibitem[{{Heckathorn}, {Bruhweiler} \& {Gull}(1982){Heckathorn}, {Bruhweiler},
  \& {Gull}}]{hec82}
{Heckathorn} J.~N., {Bruhweiler} F.~C., {Gull} T.~R., 1982, \apj, 252, 230

\bibitem[{{Heyer} {et~al}\mbox{.}(1998){Heyer}, {Brunt}, {Snell}, {Howe},
  {Schloerb}, \& {Carpenter}}]{hey98}
{Heyer} M.~H., {Brunt} C., {Snell} R.~L., {Howe} J.~E., {Schloerb} F.~P.,
  {Carpenter} J.~M., 1998, \apjs, 115, 241

\bibitem[{{Higuchi} {et~al}\mbox{.}(2013){Higuchi}, {Kurono}, {Naoi}, {Saito},
  {Mauersberger}, \& {Kawabe}}]{hig13}
{Higuchi} A.~E., {Kurono} Y., {Naoi} T., {Saito} M., {Mauersberger} R.,
  {Kawabe} R., 2013, \apj, 765, 101

\bibitem[{{Jaschek} \& {Egret}(1982)}]{jas82}
{Jaschek} M., {Egret} D., 1982, in IAU Symposium, Vol.~98, Be Stars, {Jaschek}
  M., {Groth} H.-G., eds., p. 261

\bibitem[{{Junkes}, {Fuerst} \& {Reich}(1992){Junkes}, {Fuerst}, \&
  {Reich}}]{jun92}
{Junkes} N., {Fuerst} E., {Reich} W., 1992, \aap, 261, 289

\bibitem[{{Kerton} \& {Martin}(2000)}]{ker00}
{Kerton} C.~R., {Martin} P.~G., 2000, \apjs, 126, 85

\bibitem[{{Koenig} {et~al}\mbox{.}(2012){Koenig}, {Leisawitz}, {Benford},
  {Rebull}, {Padgett}, \& {Assef}}]{koe12}
{Koenig} X.~P., {Leisawitz} D.~T., {Benford} D.~J., {Rebull} L.~M., {Padgett}
  D.~L., {Assef} R.~J., 2012, \apj, 744, 130

\bibitem[{{Kohoutek} \& {Wehmeyer}(1997)}]{koh97}
{Kohoutek} L., {Wehmeyer} R., 1997, Astronomische Abhandlungen der Hamburger
  Sternwarte, 11

\bibitem[{{Koo} \& {McKee}(1992)}]{koo92}
{Koo} B.-C., {McKee} C.~F., 1992, \apj, 388, 103

\bibitem[{{Lefloch} \& {Lazareff}(1994)}]{lef94}
{Lefloch} B., {Lazareff} B., 1994, \aap, 289, 559

\bibitem[{{Liu}, {Wu} \& {Zhang}(2013){Liu}, {Wu}, \& {Zhang}}]{liu13}
{Liu} T., {Wu} Y., {Zhang} H., 2013, in IAU Symposium, Vol. 292, IAU Symposium,
  {Wong} T., {Ott} J., eds., pp. 48--48

\bibitem[{{Liu} {et~al}\mbox{.}(2012){Liu}, {Wu}, {Zhang}, \& {Qin}}]{liu12}
{Liu} T., {Wu} Y., {Zhang} H., {Qin} S.-L., 2012, \apj, 751, 68

\bibitem[{{Lumsden} {et~al}\mbox{.}(2002){Lumsden}, {Hoare}, {Oudmaijer}, \&
  {Richards}}]{lum02}
{Lumsden} S.~L., {Hoare} M.~G., {Oudmaijer} R.~D., {Richards} D., 2002, \mnras,
  336, 621

\bibitem[{{Ma{\'{\i}}z Apell{\'a}niz} {et~al}\mbox{.}(2013){Ma{\'{\i}}z
  Apell{\'a}niz}, {Sota}, {Morrell}, {Barb{\'a}}, {Walborn}, {Alfaro}, {Gamen},
  {Arias}, \& {Gallego Calvente}}]{mai13}
{Ma{\'{\i}}z Apell{\'a}niz} J. {et~al.}, 2013, in Massive Stars: From alpha to
  Omega, p. 198

\bibitem[{{Mathis}, {Mezger} \& {Panagia}(1983){Mathis}, {Mezger}, \&
  {Panagia}}]{mat83}
{Mathis} J.~S., {Mezger} P.~G., {Panagia} N., 1983, \aap, 128, 212

\bibitem[{{Mezger} \& {Henderson}(1967)}]{mez67}
{Mezger} P.~G., {Henderson} A.~P., 1967, \apj, 147, 471

\bibitem[{{Molinari} {et~al}\mbox{.}(2008){Molinari}, {Pezzuto}, {Cesaroni},
  {Brand}, {Faustini}, \& {Testi}}]{mol08}
{Molinari} S., {Pezzuto} S., {Cesaroni} R., {Brand} J., {Faustini} F., {Testi}
  L., 2008, \aap, 481, 345

\bibitem[{{Molinari} {et~al}\mbox{.}(2011){Molinari}, {Schisano}, {Faustini},
  {Pestalozzi}, {di Giorgio}, \& {Liu}}]{mol11}
{Molinari} S., {Schisano} E., {Faustini} F., {Pestalozzi} M., {di Giorgio}
  A.~M., {Liu} S., 2011, \aap, 530, A133

\bibitem[{{Molinari} {et~al}\mbox{.}(2010){Molinari}, {Swinyard}, {Bally},
  {Barlow}, {Bernard}, {Martin}, {Moore}, {Noriega-Crespo}, {Plume}, {Testi},
  {Zavagno}, {Abergel}, {Ali}, {Andr{\'e}}, {Baluteau}, {Benedettini},
  {Bern{\'e}}, {Billot}, {Blommaert}, {Bontemps}, {Boulanger}, {Brand},
  {Brunt}, {Burton}, {Campeggio}, {Carey}, {Caselli}, {Cesaroni}, {Cernicharo},
  {Chakrabarti}, {Chrysostomou}, {Codella}, {Cohen}, {Compiegne}, {Davis}, {de
  Bernardis}, {de Gasperis}, {Di Francesco}, {di Giorgio}, {Elia}, {Faustini},
  {Fischera}, {Fukui}, {Fuller}, {Ganga}, {Garcia-Lario}, {Giard}, {Giardino},
  {Glenn}, {Goldsmith}, {Griffin}, {Hoare}, {Huang}, {Jiang}, {Joblin},
  {Joncas}, {Juvela}, {Kirk}, {Lagache}, {Li}, {Lim}, {Lord}, {Lucas},
  {Maiolo}, {Marengo}, {Marshall}, {Masi}, {Massi}, {Matsuura}, {Meny},
  {Minier}, {Miville-Desch{\^e}nes}, {Montier}, {Motte}, {M{\"u}ller},
  {Natoli}, {Neves}, {Olmi}, {Paladini}, {Paradis}, {Pestalozzi}, {Pezzuto},
  {Piacentini}, {Pomar{\`e}s}, {Popescu}, {Reach}, {Richer}, {Ristorcelli},
  {Roy}, {Royer}, {Russeil}, {Saraceno}, {Sauvage}, {Schilke},
  {Schneider-Bontemps}, {Schuller}, {Schultz}, {Shepherd}, {Sibthorpe},
  {Smith}, {Smith}, {Spinoglio}, {Stamatellos}, {Strafella}, {Stringfellow},
  {Sturm}, {Taylor}, {Thompson}, {Tuffs}, {Umana}, {Valenziano}, {Vavrek},
  {Viti}, {Waelkens}, {Ward-Thompson}, {White}, {Wyrowski}, {Yorke}, \&
  {Zhang}}]{mol10}
{Molinari} S. {et~al.}, 2010, \pasp, 122, 314

\bibitem[{{Morgan}, {Urquhart} \& {Thompson}(2009){Morgan}, {Urquhart}, \&
  {Thompson}}]{mor09}
{Morgan} L.~K., {Urquhart} J.~S., {Thompson} M.~A., 2009, \mnras, 400, 1726

\bibitem[{{Nakanishi} \& {Sofue}(2006)}]{nak06}
{Nakanishi} H., {Sofue} Y., 2006, \pasj, 58, 847

\bibitem[{{Paladini} {et~al}\mbox{.}(2012){Paladini}, {Umana}, {Veneziani},
  {Noriega-Crespo}, {Anderson}, {Piacentini}, {Pinheiro Gon{\c c}alves},
  {Paradis}, {Tibbs}, {Bernard}, \& {Natoli}}]{pal12}
{Paladini} R. {et~al.}, 2012, \apj, 760, 149

\bibitem[{{Piazzo} {et~al}\mbox{.}(2015){Piazzo}, {Calzoletti}, {Faustini},
  {Pestalozzi}, {Pezzuto}, {Elia}, {di Giorgio}, \& {Molinari}}]{pia15}
{Piazzo} L., {Calzoletti} L., {Faustini} F., {Pestalozzi} M., {Pezzuto} S.,
  {Elia} D., {di Giorgio} A., {Molinari} S., 2015, \mnras, 447, 1471

\bibitem[{{Pilbratt} {et~al}\mbox{.}(2010){Pilbratt}, {Riedinger}, {Passvogel},
  {Crone}, {Doyle}, {Gageur}, {Heras}, {Jewell}, {Metcalfe}, {Ott}, \&
  {Schmidt}}]{pil10}
{Pilbratt} G.~L. {et~al.}, 2010, \aap, 518, L1

\bibitem[{{Pineault}(1998)}]{pin98}
{Pineault} S., 1998, \aj, 115, 2483

\bibitem[{{Poglitsch} {et~al}\mbox{.}(2010){Poglitsch}, {Waelkens}, {Geis},
  {Feuchtgruber}, {Vandenbussche}, {Rodriguez}, {Krause}, {Renotte}, {van
  Hoof}, {Saraceno}, {Cepa}, {Kerschbaum}, {Agn{\`e}se}, {Ali}, {Altieri},
  {Andreani}, {Augueres}, {Balog}, {Barl}, {Bauer}, {Belbachir}, {Benedettini},
  {Billot}, {Boulade}, {Bischof}, {Blommaert}, {Callut}, {Cara}, {Cerulli},
  {Cesarsky}, {Contursi}, {Creten}, {De Meester}, {Doublier}, {Doumayrou},
  {Duband}, {Exter}, {Genzel}, {Gillis}, {Gr{\"o}zinger}, {Henning},
  {Herreros}, {Huygen}, {Inguscio}, {Jakob}, {Jamar}, {Jean}, {de Jong},
  {Katterloher}, {Kiss}, {Klaas}, {Lemke}, {Lutz}, {Madden}, {Marquet},
  {Martignac}, {Mazy}, {Merken}, {Montfort}, {Morbidelli}, {M{\"u}ller},
  {Nielbock}, {Okumura}, {Orfei}, {Ottensamer}, {Pezzuto}, {Popesso},
  {Putzeys}, {Regibo}, {Reveret}, {Royer}, {Sauvage}, {Schreiber}, {Stegmaier},
  {Schmitt}, {Schubert}, {Sturm}, {Thiel}, {Tofani}, {Vavrek}, {Wetzstein},
  {Wieprecht}, \& {Wiezorrek}}]{pog10}
{Poglitsch} A. {et~al.}, 2010, \aap, 518, L2

\bibitem[{{Reed}(2003)}]{ree03}
{Reed} B.~C., 2003, \aj, 125, 2531

\bibitem[{{Robitaille} \& {Whitney}(2014)}]{rob14}
{Robitaille} T.~P., {Whitney} B.~A., 2014, Astrophysics and Space Science
  Proceedings, 36, 157

\bibitem[{{Robitaille} {et~al}\mbox{.}(2007){Robitaille}, {Whitney},
  {Indebetouw}, \& {Wood}}]{rob07}
{Robitaille} T.~P., {Whitney} B.~A., {Indebetouw} R., {Wood} K., 2007, \apjs,
  169, 328

\bibitem[{{Robitaille} {et~al}\mbox{.}(2006){Robitaille}, {Whitney},
  {Indebetouw}, {Wood}, \& {Denzmore}}]{rob06}
{Robitaille} T.~P., {Whitney} B.~A., {Indebetouw} R., {Wood} K., {Denzmore} P.,
  2006, \apjs, 167, 256

\bibitem[{{Rod{\'o}n} {et~al}\mbox{.}(2010){Rod{\'o}n}, {Zavagno}, {Baluteau},
  {Anderson}, {Polehampton}, {Abergel}, {Motte}, {Bontemps}, {Ade},
  {Andr{\'e}}, {Arab}, {Beichman}, {Bernard}, {Blagrave}, {Boulanger}, {Cohen},
  {Compiegne}, {Cox}, {Dartois}, {Davis}, {Emery}, {Fulton}, {Gry}, {Habart},
  {Halpern}, {Huang}, {Joblin}, {Jones}, {Kirk}, {Lagache}, {Lin}, {Madden},
  {Makiwa}, {Martin}, {Miville-Desch{\^e}nes}, {Molinari}, {Moseley}, {Naylor},
  {Okumura}, {Orieux}, {Pinheiro Gon{\c c}alves}, {Rodet}, {Russeil},
  {Saraceno}, {Sidher}, {Spencer}, {Swinyard}, {Ward-Thompson}, \&
  {White}}]{rod10}
{Rod{\'o}n} J.~A. {et~al.}, 2010, \aap, 518, L80

\bibitem[{{Schwartz}(1985)}]{sch85}
{Schwartz} P.~R., 1985, \apj, 298, 292

\bibitem[{{Smith} \& {Conti}(2008)}]{smi08}
{Smith} N., {Conti} P.~S., 2008, \apj, 679, 1467

\bibitem[{{Stock}, {Nassau} \& {Stephenson}(1960){Stock}, {Nassau}, \&
  {Stephenson}}]{sto60}
{Stock} J., {Nassau} J.~J., {Stephenson} C.~B., 1960, Hamburger Sternw.~Warner
  \& Swasey Obs., 0

\bibitem[{{Tan} {et~al}\mbox{.}(2014){Tan}, {Beltr{\'a}n}, {Caselli},
  {Fontani}, {Fuente}, {Krumholz}, {McKee}, \& {Stolte}}]{tan14}
{Tan} J.~C., {Beltr{\'a}n} M.~T., {Caselli} P., {Fontani} F., {Fuente} A.,
  {Krumholz} M.~R., {McKee} C.~F., {Stolte} A., 2014, Protostars and Planets
  VI, 149

\bibitem[{{Taylor} {et~al}\mbox{.}(2003){Taylor}, {Gibson}, {Peracaula},
  {Martin}, {Landecker}, {Brunt}, {Dewdney}, {Dougherty}, {Gray}, {Higgs},
  {Kerton}, {Knee}, {Kothes}, {Purton}, {Uyaniker}, {Wallace}, {Willis}, \&
  {Durand}}]{tay03}
{Taylor} A.~R. {et~al.}, 2003, \aj, 125, 3145

\bibitem[{{Traficante} {et~al}\mbox{.}(2011){Traficante}, {Calzoletti},
  {Veneziani}, {Ali}, {de Gasperis}, {di Giorgio}, {Faustini}, {Ikhenaode},
  {Molinari}, {Natoli}, {Pestalozzi}, {Pezzuto}, {Piacentini}, {Piazzo},
  {Polenta}, \& {Schisano}}]{tra11}
{Traficante} A. {et~al.}, 2011, \mnras, 416, 2932

\bibitem[{{Tremblin} {et~al}\mbox{.}(2012{\natexlab{a}}){Tremblin}, {Audit},
  {Minier}, {Schmidt}, \& {Schneider}}]{tre12a}
{Tremblin} P., {Audit} E., {Minier} V., {Schmidt} W., {Schneider} N.,
  2012{\natexlab{a}}, \aap

\bibitem[{{Tremblin} {et~al}\mbox{.}(2012{\natexlab{b}}){Tremblin}, {Audit},
  {Minier}, \& {Schneider}}]{tre12b}
{Tremblin} P., {Audit} E., {Minier} V., {Schneider} N., 2012{\natexlab{b}},
  \aap

\bibitem[{{Tremblin} {et~al}\mbox{.}(2013){Tremblin}, {Minier}, {Schneider},
  {Audit}, {Hill}, {Didelon}, {Peretto}, {Arzoumanian}, {Motte}, {Zavagno},
  {Bontemps}, {Anderson}, {Andr{\'e}}, {Bernard}, {Csengeri}, {Di Francesco},
  {Elia}, {Hennemann}, {K{\"o}nyves}, {Marston}, {Nguyen Luong},
  {Rivera-Ingraham}, {Roussel}, {Sousbie}, {Spinoglio}, {White}, \&
  {Williams}}]{tre13}
{Tremblin} P. {et~al.}, 2013, \aap, 560, A19

\bibitem[{{Urquhart} {et~al}\mbox{.}(2008){Urquhart}, {Hoare}, {Lumsden},
  {Oudmaijer}, \& {Moore}}]{ur208}
{Urquhart} J.~S., {Hoare} M.~G., {Lumsden} S.~L., {Oudmaijer} R.~D., {Moore}
  T.~J.~T., 2008, in Astronomical Society of the Pacific Conference Series,
  Vol. 387, Massive Star Formation: Observations Confront Theory, {Beuther} H.,
  {Linz} H., {Henning} T., eds., p. 381

\bibitem[{{Urquhart} {et~al}\mbox{.}(2009){Urquhart}, {Hoare}, {Purcell},
  {Lumsden}, {Oudmaijer}, {Moore}, {Busfield}, {Mottram}, \& {Davies}}]{urq09}
{Urquhart} J.~S. {et~al.}, 2009, \aap, 501, 539

\bibitem[{{Urquhart} {et~al}\mbox{.}(2011){Urquhart}, {Morgan}, {Figura},
  {Moore}, {Lumsden}, {Hoare}, {Oudmaijer}, {Mottram}, {Davies}, \&
  {Dunham}}]{urq11}
{Urquhart} J.~S. {et~al.}, 2011, \mnras, 418, 1689

\bibitem[{{van der Hucht}(2001)}]{van01}
{van der Hucht} K.~A., 2001, \nar, 45, 135

\bibitem[{{Wackerling}(1970)}]{wac70}
{Wackerling} L.~R., 1970, \mnras, 73, 153

\bibitem[{{Weaver} {et~al}\mbox{.}(1977){Weaver}, {McCray}, {Castor},
  {Shapiro}, \& {Moore}}]{wea77}
{Weaver} R., {McCray} R., {Castor} J., {Shapiro} P., {Moore} R., 1977, \apj,
  218, 377

\bibitem[{{Wright} {et~al}\mbox{.}(2010){Wright}, {Eisenhardt}, {Mainzer},
  {Ressler}, {Cutri}, {Jarrett}, {Kirkpatrick}, {Padgett}, {McMillan},
  {Skrutskie}, {Stanford}, {Cohen}, {Walker}, {Mather}, {Leisawitz}, {Gautier},
  {McLean}, {Benford}, {Lonsdale}, {Blain}, {Mendez}, {Irace}, {Duval}, {Liu},
  {Royer}, {Heinrichsen}, {Howard}, {Shannon}, {Kendall}, {Walsh}, {Larsen},
  {Cardon}, {Schick}, {Schwalm}, {Abid}, {Fabinsky}, {Naes}, \& {Tsai}}]{wri10}
{Wright} E.~L. {et~al.}, 2010, \aj, 140, 1868

\bibitem[{{Zavagno} {et~al}\mbox{.}(2010){Zavagno}, {Russeil}, {Motte},
  {Anderson}, {Deharveng}, {Rod{\'o}n}, {Bontemps}, {Abergel}, {Baluteau},
  {Sauvage}, {Andr{\'e}}, {Hill}, \& {White}}]{zav10}
{Zavagno} A. {et~al.}, 2010, \aap, 518, L81

\end{thebibliography}

\IfFileExists{\jobname.bbl}{}
{\typeout{}
\typeout{****************************************************}
\typeout{****************************************************}
\typeout{** Please run "bibtex \jobname" to optain}
\typeout{** the bibliography and then re-run LaTeX}
\typeout{** twice to fix the references!}
\typeout{****************************************************}
\typeout{****************************************************}
\typeout{}
}

\end{document}